\documentclass[twocolumn]{aastex62}
\usepackage{graphicx}
\usepackage{dcolumn}
\usepackage{bm}
\usepackage{natbib}

\begin{document} 

\title{New Neutron-Capture Site in Massive Pop III and Pop II Stars as a Source for Heavy Elements in the Early Galaxy} 
\correspondingauthor{Projjwal Banerjee}
\email{projjwal@sjtu.edu.cn}
\author{Projjwal Banerjee}
\affil{Department of Astronomy,  School of Physics and Astronomy, Shanghai Jiao Tong University, Shanghai 200240, China}
\author{Yong-Zhong Qian}
\affil{School of Physics and Astronomy, University of Minnesota, Minneapolis, MN 55455}
\affil{Tsung-Dao Lee Institute, Shanghai 200240, China}
\author{Alexander Heger}
\affil{Monash Centre for Astrophysics, School of Physics and Astronomy,
Monash University, Vic 3800, Australia}
\affil{Tsung-Dao Lee Institute, Shanghai 200240, China}

\date{\today}

\newcommand{\A}[1]{\textcolor{green}{\textbf{#1}}}

\begin{abstract}
We propose a new neutron-capture site in  
early metal-poor and metal-free stars of $\sim 20$--$30 \,\mathrm{M}_\odot$ 
that results from proton ingestion in the He shell during late stages of the 
stars' lives. Most of the neutron capture occurs in the first $\lesssim 10^6\,$s
following proton ingestion when $^{13}{\rm C}(\alpha,\mathrm{n})^{16}\mathrm{O}$ produces
neutron densities typical of the intermediate neutron-capture process. 
This phase may be followed by another lasting $\gtrsim 10^7\,$s 
with $^{17}\mathrm{O}(\alpha,\mathrm{n})^{20}\mathrm{Ne}$ producing much lower neutron densities
typical of the slow neutron-capture process. 
We explore the dependence of the proposed neutron-capture nucleosynthesis 
on the amount and time of proton ingestion, 
the initial metallicity, and the ensuing supernova shock.
We obtain a range of heavy element abundance patterns
including those attributed to the slow neutron-capture process or
a combination of the slow and rapid neutron-capture processes. Our results can 
account for the observed ubiquity of heavy elements such as Sr and Ba in the early Galaxy and explain
puzzling abundance patterns of these elements in at least some very metal-poor (VMP) stars including 
those of the carbon-enhanced varieties.
In the latter case, the explanation by the single site proposed here differs from
the existing paradigm that attributes 
various classes of VMP stars to enrichment by multiple different sites. 
\end{abstract}

\section{Introduction}
Massive stars of $\gtrsim 8 \,\mathrm{M}_\odot$ have lifetimes of $\sim 10\,$Myr and die as core-collapse 
supernovae (CCSNe). They are the predominant source for chemical enrichment of the interstellar 
medium (ISM) during the first $\sim 1\,$Gyr after the Big Bang. Very metal-poor (VMP) low-mass 
stars formed at those early times have typical Fe abundances less than about
one-hundredth the solar value ($[\textrm{Fe}/\textrm{H}]\equiv\log(\mathrm{Fe}/\mathrm{H})-\log(\mathrm{Fe}/\mathrm{H})_\odot\lesssim-2$).
They live until today and are fossil records of the early ISM. 
Elements heavier than the Fe group like Sr (mass number $A\sim 88$)  and Ba ($A\sim 135$--$138$), 
whose solar abundances were mainly produced by the slow (\textsl{s}) and rapid (\textsl{r}) 
neutron-capture processes, have been observed in the majority of VMP stars \citep{roederer}. 
Moreover, measurements of multiple heavy elements in individual stars have 
revealed diverse abundance patterns encompassing those characteristic of the \textsl{r} or \textsl{s} process,
and those in between for the so-called \textsl{r}/\textsl{s} stars \citep{beers2005}. 
The ubiquity of heavy elements and the diversity of their patterns in VMP stars, however, 
are difficult to explain in the existing framework. Regular CCSNe can only produce elements 
up to $A\lesssim 120$ in the neutrino-driven wind \citep{roberts,arcones}, and therefore,
can only account for the presence of Sr but not the ubiquity of heavier elements like Ba.
Although rare events associated with massive stars, such as jet-induced CCSNe 
\citep{winteler2012} or neutron star mergers \citep{lattimer1974,eichler1989,rosswog1999}, 
can produce Ba by the \textsl{r}-process, they are too infrequent to have provided 
pervasive enrichment to the ISM and cannot explain patterns sharply different from
the \textsl{r}-process kind. This is consistent with only $\sim3$--$5\,\%$ of the VMP stars having 
strong \textsl{r}-process enrichment \citep{roederer2014b}.

Particularly puzzling are the carbon enhanced metal-poor (CEMP) stars 
with $[\textrm{C}/\textrm{Fe}]> 0.7$, which are further divided into three groups depending on their 
heavy element enrichment \citep{beers2005}. 
Stars with low levels of heavy elements ($[\textrm{Ba}/\textrm{Fe}]< 0$) are called 
CEMP-no stars. These are currently thought to have formed from an ISM polluted by 
massive stars of the very first (Pop III) and very early (Pop II) generations. Several models 
have been put forward for the polluting sources, which include rotating massive stars \citep{meynet2006}
and CCSNe with low to medium explosion energies. The latter source preferentially 
ejected C relative to Fe, leading to $[\textrm{C}/\textrm{Fe}]> 0.7$ 
\citep{umeda2005,nomoto2006,tominaga2014,hw2010}.
The origin of the heavy elements such as Ba in the CEMP-no stars, especially those thought to have been 
enriched by Pop III sources, remains unclear.

Another subclass of CEMP stars is the so-called CEMP-\textsl{s} stars with
\textsl{s}-process-like patterns ($[\textrm{Ba}/\textrm{Eu}]> 0.5$) and high enrichment of heavy elements ($[\textrm{Ba}/\textrm{Fe}]> 1$)
\citep{beers2005}. In contrast to CEMP-no stars, these are thought to be the result of 
surface pollution by binary companions and do not reflect the composition of the ISM from which they were 
born. Specifically, the primary member of a binary
produced the \textsl{s}-process elements during its asymptotic giant branch (AGB) phase 
and then transferred these elements along with C to the secondary member, which 
is observed as a CEMP-\textsl{s} star today. This prevailing scenario can 
explain the abundance patterns of heavy elements in most CEMP-\textsl{s} stars reasonably well, 
although patterns in some CEMP-\textsl{s} stars remain a challenge for this mechanism \citep{bisterzo3}. 
In addition, whereas most CEMP-\textsl{s} appear to be in binaries \citep{lucatello2005}
as required by the above binary mass transfer scenario,  
recent observations indicate that $\sim 10$--$30\,\%$ of such stars could be single \citep{hansenCEMPs}. 
Therefore, some other mechanism is required to provide \textsl{s}-process enrichment to single CEMP-\textsl{s} stars. 
Fast-rotating metal-poor massive stars (``spinstars'') have been proposed as
another \textsl{s}-process site in the early Galaxy \citep{pignatari2008,frischk2016}, but
they dominantly produce elements around Sr with some Ba and very little Pb ($A\sim 208$). 
Therefore, they cannot explain the overall patterns in those CEMP-\textsl{s} stars with substantial Pb abundances.

The so-called CEMP-\textsl{r}/\textsl{s} stars form yet another subclass that shows high 
enrichment in heavy elements with patterns in between those for the \textsl{r} and 
\textsl{s} processes ($0<[\textrm{Ba}/\textrm{Eu}]<0.5$). The origin of these stars remains a major
puzzle although various scenarios have been proposed
\citep{cohen2003,barbuy2005,jonsell2006}. In a popular scenario, 
CEMP-\textsl{r}/\textsl{s} stars have a similar origin to CEMP-\textsl{s} stars in binaries, 
except that the former stars were born in highly \textsl{r}-process-enriched
parent clouds, thereby acquiring their \textsl{r}-like features. 
Their \textsl{s}-like features were due to binary mass transfer just like the CEMP-\textsl{s} stars.
Recent studies, however, have shown 
that this scenario is disfavored and that
most likely a new neutron-capture site 
produced both the \textsl{r} and \textsl{s}-like features 
in CEMP-\textsl{r}/\textsl{s} stars \citep{lugaro2012}. 
The so-called intermediate (\textsl{i}) process \citep{cowan1977} has been proposed as a possible mechanism. 
This process operates at 
neutron densities much higher than those typical of the \textsl{s}-process but 
lower than those for the \textsl{r}-process.
It has been shown that
the required \textsl{i}-process neutron densities of 
up to $\sim 10^{15}\,\mathrm{cm}^{-3}$ can be
generated via $^{12}\mathrm{C}(\mathrm{p},\gamma)^{13}{\rm N}(e^+\nu_e)^{13}\mathrm{C}(\alpha,\mathrm{n})^{16}\mathrm{O}$
following proton ingestion in He layers of low to intermediate mass stars during late stages of their evolution 
\citep{fujimoto2000,campbell2010,herwig2011,jones2016}. Transfer of the \textsl{i}-process products
along with C to a low mass binary companion  
would result in the latter being observed as a CEMP-\textsl{r}/\textsl{s} star today.  
One-zone parametric studies by \citet{dardelet2014} and more recently by \citet{hampel2016} 
have shown that the \textsl{i}-process can produce the abundance patterns of heavy elements in many CEMP-\textsl{r}/\textsl{s} stars very well. 
Therefore, the leading scenarios for both CEMP-\textsl{s} and
CEMP-\textsl{r}/\textsl{s} stars require mass transfer following neutron-capture nucleosynthesis by
low to intermediate mass companions in binaries. We note that rapidly accreting white dwarfs were proposed as another
site for the \textsl{i}-process \citep{deni2017}.

Here we report a new site for heavy element synthesis by neutron capture
in early massive stars of $\sim 20$--$30\,\mathrm{M}_\odot$ with zero (Pop III) to low (Pop II) 
metallicity ($[\textrm{Fe}/\textrm{H}]\lesssim -2$) at birth. A few years prior to the end of such a star's life,
C has been depleted in the center and the He shell becomes convective. Protons are 
present at low levels in the outer He shell and some of them are ingested into the inner
He shell by convective boundary mixing. Further transport to the hotter region initiates the 
familiar reaction sequence
$^{12}{\rm C}(\mathrm{p},\gamma)^{13}{\rm N}(e^+\nu_e)^{13}{\rm C}(\alpha,\mathrm{n})^{16}{\rm O}$ that produces 
neutron densities corresponding to both the \textsl{i}-process and the 
\textsl{s}-process. As a result, elements up to Bi ($A=209$) are produced. The final abundance pattern can vary from
\textsl{s}-like to \textsl{r}/\textsl{s}-like, depending on the amount of proton ingestion and the time 
available for neutron capture before core collapse. Instances of proton 
ingestion in the convective He shell have been reported earlier for
zero-metallicity stars with similar masses to the above range \citep{hw2010,limongi2012}. 
After we completed the study for this paper, \citet{clarkson2017} reported work on
proton ingestion in the convective He shell in a zero-metallicity $45 \,\mathrm{M}_\odot$ star, which led to 
large energy generation and neutron capture. That ``\textsl{i}-process'', however, produced elements only up to the Fe group.
Here we present a detailed analysis of the neutron-capture nucleosynthesis in Pop III and Pop II stars of
$\sim 20$--$30\,\mathrm{M}_\odot$, focusing on a $25 \,\mathrm{M}_\odot$ star with an initial metallicity of zero to $[Z]=-1$.
Specifically, we examine the dependence of the nucleosynthesis on the amount and time of proton ingestion,
among other things.

The neutron-capture site presented in this study resides in a significant fraction of early 
massive stars and can explain a number of puzzles:
\begin{enumerate}
\item the ubiquity of heavy elements such as Ba in VMP stars, including CEMP-no stars with 
low enhancement of such elements, some of which
are considered as records of nucleosynthesis by the first generation of stars,

\item heavy-element patterns in some CEMP-\textsl{s} stars, including single stars whose
enrichment cannot be explained by binary mass transfer,

\item the origin of heavy elements in some CEMP-\textsl{r}/\textsl{s} stars,

\item the origin of stars with heavy-element patterns similar to CEMP-\textsl{s} and CEMP-\textsl{r}/\textsl{s} stars but 
with much lower enhancement \citep{spite2014}, and 

\item the early onset of \textsl{s}-like signature 
in VMP stars \citep{simmerer,saga}.
\end{enumerate}

We discuss our methods of modeling proton ingestion and the associated 
nucleosynthesis in early massive stars in \S\ref{sec-method}. 
The results on nucleosynthesis are presented and compared 
with observations in \S\ref{sec-result}. We discuss the implications of our results for 
general observations of VMP stars in \S\ref{sec-discuss}. We summarize and give our outlook in \S\ref{sec-sum}.

\section{Methods}
\label{sec-method}
\begin{figure}[h]
\centerline{\includegraphics[width=85mm]{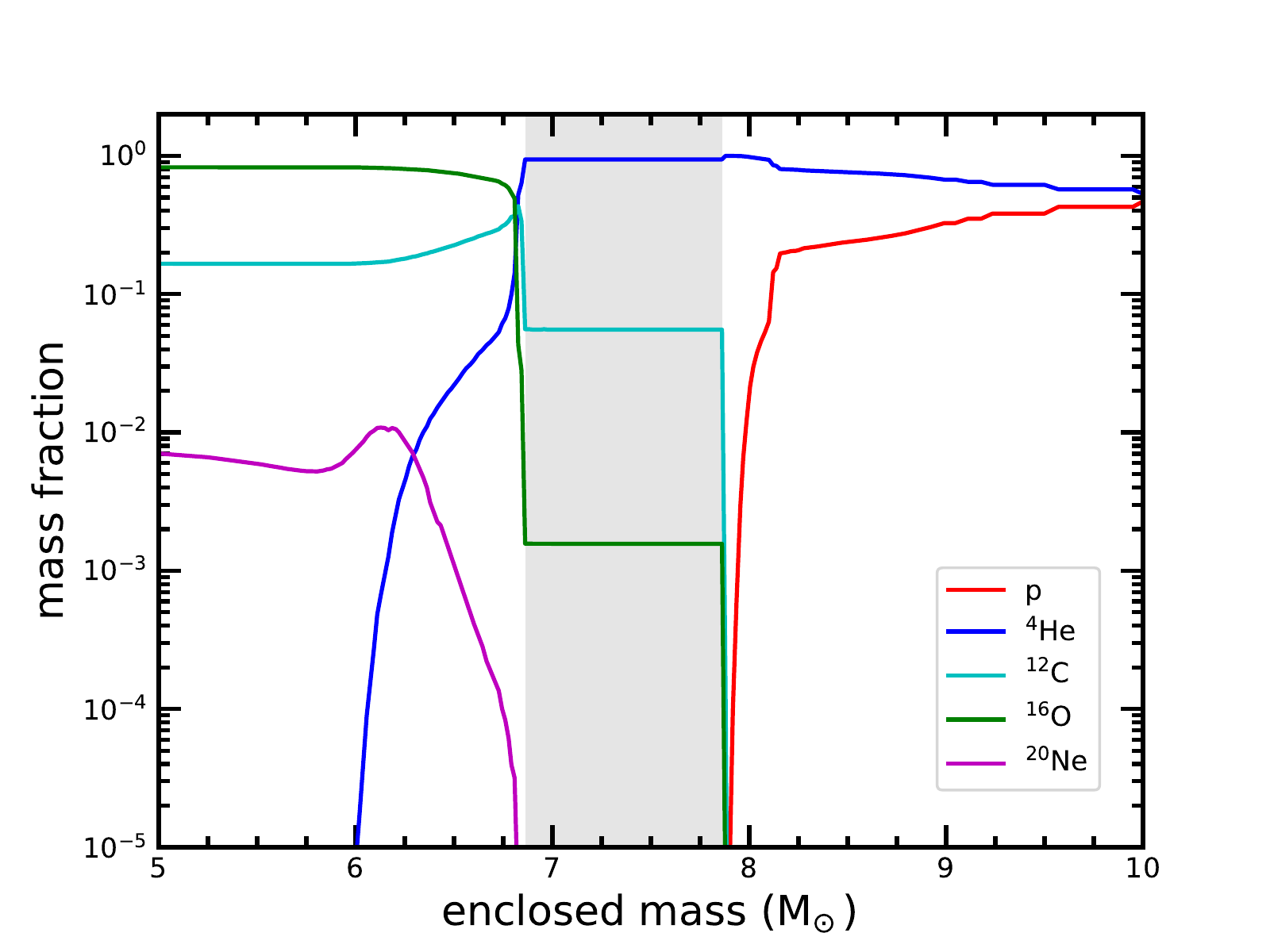}}
\caption{Mass fractions of protons (\emph{red}), $^4$He (\emph{blue}), $^{12}$C (\emph{cyan}),
$^{16}$O (\emph{green}), and $^{20}$Ne (\emph{magenta}) as functions of mass coordinate for
a metal-free $25\,\mathrm{M}_\odot$ star at central C depletion.
The convective He shell is indicated by the grey region.}
\label{fig:struc}
\end{figure}

We study nucleosynthesis in non-rotating stars of $15$--$40\,\mathrm{M}_\odot$ with
an initial metallicity of zero to $[Z]=-2$ (corresponding to $[{\rm Fe/H]}\sim -2$).
The corresponding initial composition
is scaled from the solar abundances for Be to 
Zn and taken from the results of Big Bang nucleosynthesis for H to Li.
The elements above Zn are excluded so that they can be clearly 
attributed to nucleosynthesis in the star.
We use the 1D hydrodynamic code \textsc{Kepler} \citep{weaver1978,rauscher2003} 
to calculate the evolution and associated nucleosynthesis of a star 
from its birth to its death in a core-collapse supernova (CCSN).
In all the models, we find that the base of the He shell becomes convective 
when the temperature at the center reaches $T\sim 1.2\times 10^9\,$K and most of the C there has been exhausted. 
This onset of convection occurs a few years before the CCSN as the star shrinks in order to increase 
the central temperature to burn C. For a $25 \,\mathrm{M}_\odot$ star, the radius at the base of the He shell  
decreases by $\sim 60\,\%$ from C ignition to C depletion at the center. 
As a result, the temperature at this position increases from $1.75\times 10^8\,$K
to $\sim 2.7 \times 10^8\,$K, causing a large increase in energy production by the triple alpha reaction. 
The excess energy generated can only be efficiently transported by convection, which is modeled with the 
standard mixing length theory (MLT) in \textsc{Kepler}. The convection at the base of the He shell grows until almost 
the entire He shell becomes fully convective when C is depleted at the center.
The structure of a metal-free $25 \,\mathrm{M}_\odot$ star at this time is shown in Fig.~\ref{fig:struc}, where 
the convective region is indicated as the gray area.

\subsection{Proton Ingestion in the Convective {\rm He} Shell}
There are $\sim 10^{-3}$--$10^{-2} \,\mathrm{M}_\odot$ of protons with mass fractions of $10^{-5}$--$10^{-2}$ 
in the outer He shell just above the boundary of the convective He shell in a $\sim 20$--$30 \,\mathrm{M}_\odot$ star.
Some of these protons can be ingested in the convective region when convection grows to encompass the 
corresponding layers or by mixing processes such as overshoot and turbulent entrainment at the convective
boundary. The default treatment of overshoot in \textsc{Kepler} allows a single zone at the boundary to mix slowly 
when a region becomes unstable to convection by the Ledoux criterion. 
Only tiny amounts of protons are ingested in a few models with this treatment.

A different overshoot prescription \citep{herwig} is used in the stellar evolution code MESA \citep{mesa}. 
In this prescription, regions outside the convective 
boundary are allowed to mix with an exponentially decaying diffusion coefficient
\begin{equation}
    D_{{\rm osht}}=D_0 \exp\left(\frac{-2z}{fH_\mathrm{P}}\right),
\end{equation}
where $D_0$ is the diffusion coefficient calculated from the MLT for a zone just inside 
the convective region, $z$ is the distance of overshoot
from the convective boundary, $H_\mathrm{P}$ is the pressure scale height, and $f$ is a dimensionless parameter
controlling the efficiency of overshoot. We modified \textsc{Kepler} to 
implement this overshoot prescription for the He shell.  
A recent study comparing \textsc{Kepler} and MESA models \citep{sukhbold} 
found that $f=0.025$ resulted in similar structure for massive stars. 
Using this number as the default value of $f$, we 
found that $\sim 10^{-5}$--$10^{-4}\,\mathrm{M}_\odot$ of protons were ingested in $\sim 20$--$30 \,\mathrm{M}_\odot$ models by the 
time of central C depletion. For models with $[Z]\lesssim -4$ or $f\gtrsim 0.025$, additional 
episodes of proton ingestion occur later when the star contracts 
during the stage from central O depletion to core collapse.

Proton ingestion does not occur in stars of $< 20\,\mathrm{M}_\odot$, where 
convection is unable to reach the layers with protons in the
outer He shell. On the other hand,
the outer He shell in stars of $> 30 \,\mathrm{M}_\odot$ has relatively high temperature and protons there 
have been burned away by the time the He shell becomes fully convective. 
So proton ingestion does not occur in such stars, either. Thus, proton ingestion is 
limited to $\sim 20$--$30 \,\mathrm{M}_\odot$ stars.

Because 1D models have large uncertainties in treating overshoot 
specifically and convective boundary mixing in general,
we used the above overshoot prescription primarily to estimate the level of proton ingestion 
that can occur in such models. Proton ingestion found in our 1D study, however, 
may also be common in 3D simulations. 
For example, additional mixing due to turbulent 
entrainment was seen in convective boundary layers in 3D stellar models \citep{arnett}. \\

\subsection{Transport of Ingested Protons in the Convective {\rm He} Shell}
In order to study the nucleosynthesis resulting from proton ingestion in detail, 
we focus on $25 \,\mathrm{M}_\odot$ stars with varying initial
metallicities. We evolve a star from its birth to the point when 
its central temperature reaches $T=1.2\times 10^9\,$K using the default mixing scheme in \textsc{Kepler}. 
Then proton ingestion is simply implemented by injecting an amount
$M_\mathrm{p}=10^{-6}$--$10^{-3} \,\mathrm{M}_\odot$ of protons
in a single zone at the upper boundary of the convective He shell at central C depletion for most models. 
Single ingestion at central O depletion and multiple episodes of ingestion are also explored 
for other models. 

Transport of ingested protons in the convective He shell is difficult to model with the standard 
MLT used in \textsc{Kepler} and most other 1D
stellar evolution codes. In fact, studies of proton ingestion in He layers of post-AGB and super-AGB stars 
\citep{herwig2011,jones2016} have shown that the MLT is inadequate. 
Similarly to these studies, we find that injecting protons at the top of the 
convection shell leads to a slightly higher entropy for the proton-enriched zone compared to 
the rest of the shell. For $M_\mathrm{p}\gtrsim 10^{-5} \,\mathrm{M}_\odot$, this entropy barrier causes a split in 
the convective He shell, where bulk of the protons are trapped in the upper 
convective zone with temperatures of $\lesssim 1.3\times 10^8\,$K.
Because of this trapping, only $\lesssim 10^{-5} \,\mathrm{M}_\odot$ of protons can reach the base of the He shell,
where the temperature is high enough to initiate neutron production via the usual reaction 
sequence $^{12}{\rm C}(\mathrm{p},\gamma)^{13}{\rm N}(e^+\nu_e)^{13}{\rm C}(\alpha,\mathrm{n})^{16}{\rm O}$.
3D simulations of AGB and post-AGB stars, however, indicate that such splitting of 
the convective He shell would likely be delayed \citep{herwig2011} or may even be 
absent \citep{stancliffe2011}. In either case, protons can travel downwards 
to hotter regions to initiate neutron production. Clearly, ingestion and 
transport of protons are intrinsically 3D phenomena and cannot be captured accurately in 1D \citep{herwig2014}.

For the reasons given above, most previous studies on 
nucleosynthesis from proton ingestion employed one-zone network
calculations using typical temperature, density, and composition 
from 1D stellar models as input. Here we perform a multi-zone calculation taking advantage of 
two separate networks incorporated in \textsc{Kepler} to compute energy generation and nucleosynthesis,
respectively. A 19 isotope network, \textsc{Approx19}, is used for energy generation and stellar structure 
evolution, whereas a large adaptive post-processing network is used to calculate nucleosynthesis 
accurately. The latter includes isotopes up to atomic number $85$ between the proton and 
neutron drip lines. 
The number of isotopes in the network grows to a maximum of $\sim 1,\!400$ 
for neutron-capture nucleosynthesis during pre-CCSN evolution, and to a maximum of $\sim 2,\!100$
for explosive nucleosynthesis during shock propagation.
The rates adopted in this study are the same as those listed in detail by \cite{rauscher2002}. 
Table~\ref{tab:rate} gives the references for the adopted rates for neutron, proton, and $\alpha$-particle 
reactions on $^{12,13}$C, $^{13,14}$N, and $^{16,17}$O. These rates are the most relevant for neutron production 
following proton ingestion. For neutron capture, theoretical rates from \citet{RATH} are used, supplemented by 
recommended experimental rates from \citet{BAAL} whenever available.
We note that most of the rates have not been measured in the astrophysically-relevant energy regime.
Whereas efforts to update all the rates are important, changes to the pertinent rates for neutron production
and most other rates are expected to be within the current uncertainties and would not qualitatively alter our results.

To avoid the issues related to splitting of the convective zone mentioned above, 
we model the ingestion of protons using only the post-processing 
network without changing the composition in \textsc{Approx19} used for energy generation. 
In the post-processing network, change in the number abundance $Y_i$ of an isotope $i$
due to convective transport and nuclear burning is modeled by
\begin{equation}
    \frac{\mathrm{d}Y_i}{\mathrm{d}t}=\left(\frac{\partial Y_i}{\partial t}\right)_{\rm nuc} + 
    \frac{\partial }{\partial M }\left[(4\pi r^2\rho)^2D_{{\rm MLT}}\frac{\partial Y_i}{\partial M}\right],
    \label{eq-dydt}
\end{equation}
where the first term on the right-hand side is due to nuclear burning 
and the second due to convective transport,
$\rho$ is the density, $M$ is the mass enclosed within radius \textsl{r}, and $D_{{\rm MLT}}$
is the MLT diffusion coefficient calculated with energy generation from \textsc{Approx19}. 
The above treatment is very similar to using the values of $D_{{\rm MLT}}$ 
before the proton ingestion for the subsequent proton transport as done in \citet{herwig2011}, where
a delayed split of the convective zone was included as an option. 
We do not attempt to model any delayed split in our calculations. 
We note that the majority of the ingested protons can travel to the bottom of the He 
shell within $\sim 10$ min in our models and any split after this time would not affect the neutron production. 

Because proton ingestion is not coupled to energy generation in our calculations, a necessary
condition for this approximate treatment is
\begin{equation}
    \epsilon_{{\rm nuc}}\tau_{{\rm conv}}\ll E_{{\rm int}},
    \label{eq-pburn}
\end{equation}
where $\epsilon_{{\rm nuc}}$ is the specific rate of energy generation from proton burning,
$\tau_{{\rm conv}}$ is the convective mixing timescale, and $E_{{\rm int}}$ is the specific internal energy 
typical of the convective region. Following \citet{jones2016}, we take
\begin{equation}
    \tau_{{\rm conv}}\approx \frac{H_\mathrm{P}}{v_{{\rm conv}}},
\end{equation}
where the pressure scale height $H_\mathrm{P}$ and the convective velocity $v_{{\rm conv}}$ are calculated from the MLT.
The post-processing network can accurately calculate the energy generation from the burning of protons
as they travel towards the hotter parts of the He shell. We have checked that the condition in
Eq.~(\ref{eq-pburn}) is satisfied for the convective He shell in our models. Take a $25 \,\mathrm{M}_\odot$ model with 
metallicty $[Z]=-3$ for example. We find that
$ \tau_{{\rm conv}}\sim (1$--$2)\times 10^4\,$s at central C depletion and
$\epsilon_{{\rm nuc}}\tau_{{\rm conv}}\lesssim 0.015\, E_{{\rm int}}$
for all regions in the convective He shell for $M_\mathrm{p}\lesssim 10^{-4} \,\mathrm{M}_\odot$. Even for $M_\mathrm{p}=10^{-3} \,\mathrm{M}_\odot$,
the highest ingested mass considered in this study, 
$\epsilon_{{\rm nuc}} \tau_{{\rm conv}}\lesssim 0.06\,E_{{\rm int}}$. 
The above discussion suggests that ignoring energy generation associated with proton ingestion 
in our models is a reasonable approximation, especially for $M_\mathrm{p}\lesssim 10^{-4} \,\mathrm{M}_\odot$.

The approximate treatment of ingestion and transport of protons described above represents
a major simplification. Nevertheless, it allows us to calculate the nucleosynthesis across multi-zones
in the convective He shell within the limitations of 1D stellar models. We prefer this treatment
to one-zone calculations although we expect qualitatively similar results for similar conditions.

\section{Results}
\label{sec-result}

The proton ingestion initiates a sequence of reactions producing neutrons, which are
captured to make heavy elements.
Below we present the results for a $25\,\mathrm{M}_\odot$ star with varying initial metallicity.
For each case, we compute the full evolution and the associated nucleosynthesis of 
the star from its birth until its death in a CCSN.
The explosion subsequent to core collapse is 
simulated by driving a piston outwards from the base of the O shell. 
The piston velocity is adjusted so that the desired explosion energy is obtained. 
The material inside the initial radius of the piston is assumed to fall back immediately onto the 
protoneutron star produced by the core collapse. 
The final yields are calculated by assuming that all the material outside this radius is ejected.

\subsection{Neutrons from Proton Ingestion and Neutron-Capture Nucleosynthesis} 

\begin{figure*}[h]
\centerline{\includegraphics[width=\textwidth]{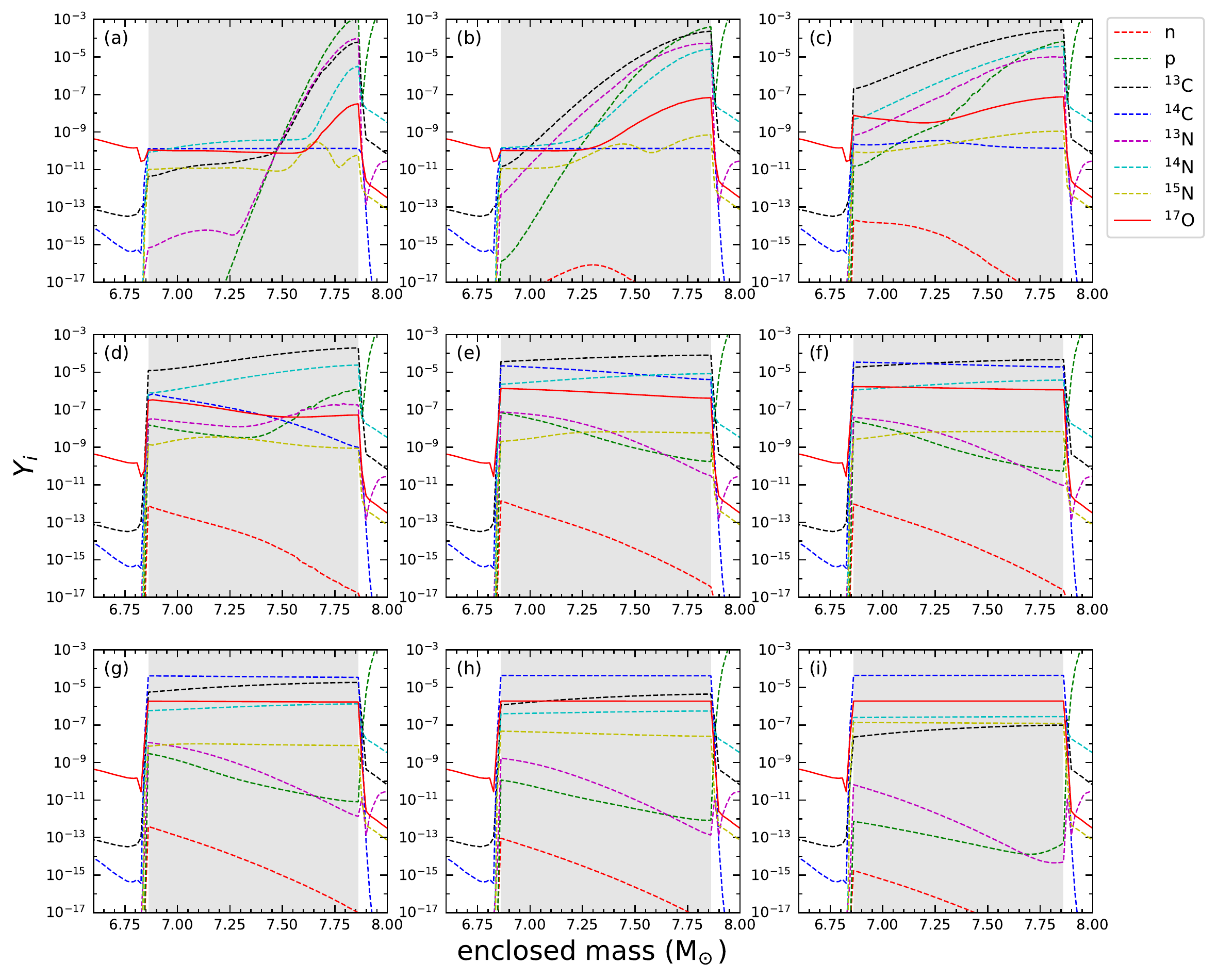}}
\caption{Abundances of neutrons (\emph{dashed red}), protons (\emph{dashed green}), $^{13}$C (\emph{dashed black}), 
$^{14}$C (\emph{dashed blue}), $^{13}$N (\emph{dashed magenta}), $^{14}$N (\emph{dashed cyan}), 
$^{15}$N (\emph{dashed yellow}), and $^{17}$O (\emph{solid red}) as functions of mass coordinate 
at different times $t$ after $10^{-4}\,\mathrm{M}_\odot$ of protons are ingested at central C depletion in a metal-free 
$25 \,\mathrm{M}_\odot$ star: 
(a) $t=8.95\times 10^2\,$s, (b) $t=3.90 \times 10^3\,$s, (c) $t=9.90 \times 10^4\,$s, (d) $t=2.39 \times 10^4\,$s, 
(e) $t=1.04 \times 10^5\,$s, (f) $t=2.04 \times 10^5\,$s, (g) $t=3.54 \times 10^5\,$s, (h) $t=5.54 \times 10^5\,$s,
and (i) $t=1.06 \times 10^6\,$s. The convective He shell is indicated by the grey region.}
\label{fig:snap}
\end{figure*}

At the top of the convective He shell where protons are ingested, the temperature
is $T\sim 10^8\,$K and the density is $\rho\sim 50\,\mathrm{g}\,\mathrm{cm}^{-3}$. 
In comparison, the conditions at the bottom of this shell are
$T\sim 2.7\times10^{8}\,$K and $\rho \sim 500\,\mathrm{g}\,\mathrm{cm}^{-3}$. The shell
also has traces of primary $^{12}$C and $^{16}$O with mass fractions of
$\sim 5\times 10^{-2}$ and $\sim 2\times10^{-3}$, respectively, produced by 
He burning. As the ingested protons are transported to the hotter region,
they are captured by $^{12}$C to produce $^{13}$N, which decays to $^{13}$C
on a time scale of $\sim 863\,$s. Neutrons are released from the subsequent reaction 
$^{13}$C$(\alpha,\mathrm{n})^{16}$O. 
The occurrence of the above processes can be clearly seen from Fig.~\ref{fig:snap}, which shows the abundance profiles of 
neutrons, protons, $^{13,14}$C, $^{13,14,15}$N, and $^{17}$O in the He shell at different times
after $10^{-4} \,\mathrm{M}_\odot$ of protons are ingested at central C depletion in a primordial $25\,\mathrm{M}_\odot$ star.
In general, the composition at a specific mass coordinate
in the He shell changes as a result of both convective mixing and nuclear reactions [see Eq.~(\ref{eq-dydt})].
Sufficiently small time steps are required to follow these changes accurately. We
find that limiting the maximum time step to $100\,$s 
is adequate to calculate the nucleosynthesis resulting
from proton ingestion.

\begin{figure*}
\centerline{\includegraphics[width=\textwidth]{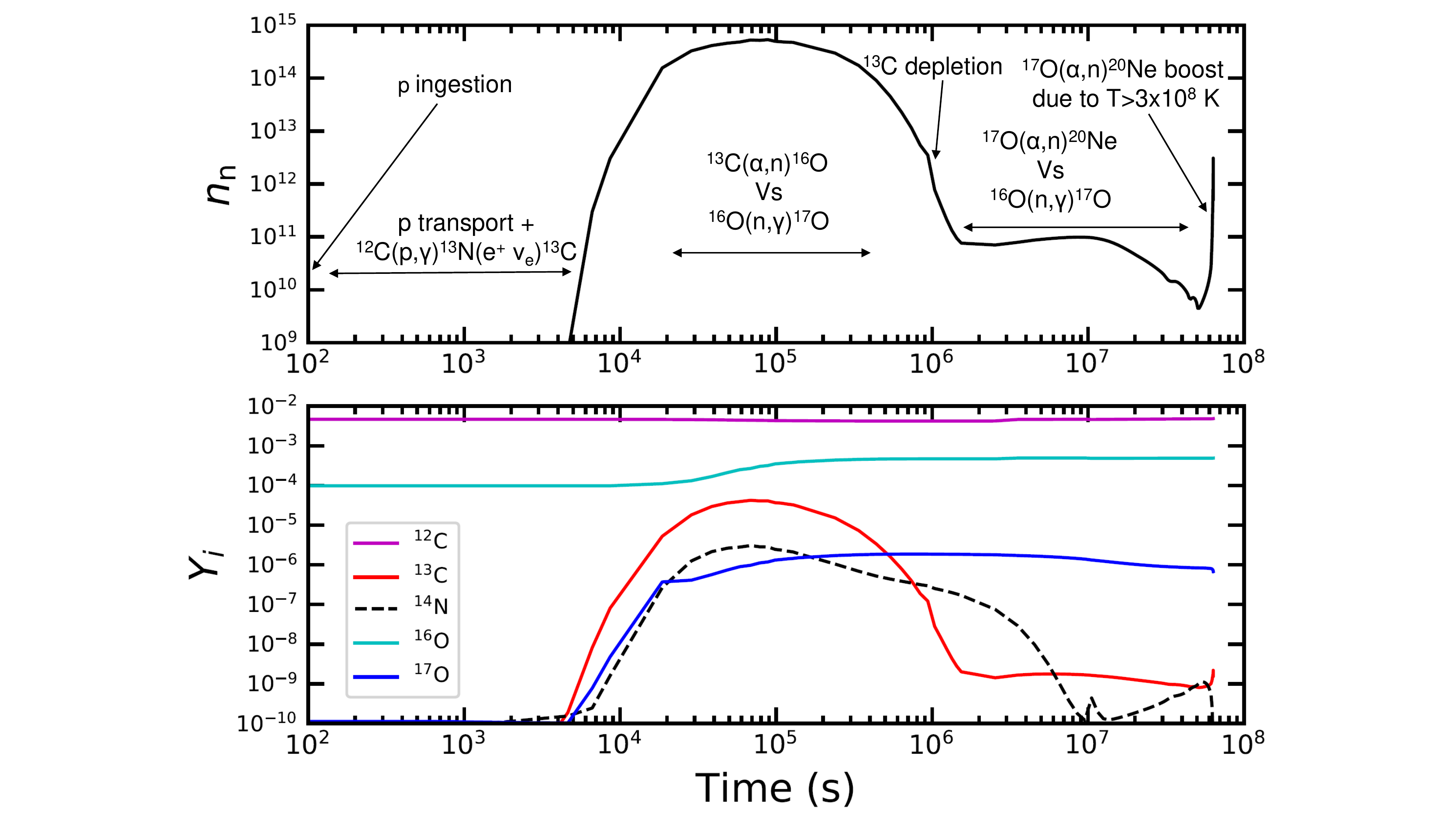}}
\caption{(Top) Neutron density $n_\mathrm{n}$ in a single zone at the base of the He shell 
of a metal-free $25\,\mathrm{M}_\odot$ star as a function of time after $10^{-4} \,\mathrm{M}_\odot$ of protons
are ingested at central C depletion. Various processes determining $n_\mathrm{n}$ are indicated.
(Bottom) Corresponding evolution of the abundance $Y_i$ for $^{12}$C (\emph{magenta}), $^{13}$C (\emph{red}), $^{14}$N (\emph{black}),
$^{16}$O (\emph{cyan}), and $^{17}$O (\emph{blue}).}
\label{fig:iso}
\end{figure*}

For illustration (see Fig.~\ref{fig:iso}), 
we describe the generic evolution of the 
neutron (number) density $n_\mathrm{n}$ in a zone (with a fixed mass coordinate) near the 
bottom of the convective He shell after $10^{-4} \,\mathrm{M}_\odot$ of protons are ingested at
the top. Transport of ingested protons to the zone 
and the associated production of $^{13}$C take $\sim 10^4\,$s. Then $n_\mathrm{n}$ sharply 
increases because neutron production by $^{13}$C$(\alpha,\mathrm{n})^{16}$O occurs 
efficiently at the relatively high temperature of the zone. The primary $^{16}$O 
from He burning is the main neutron poison through $^{16}$O$(\mathrm{n},\gamma)^{17}$O.
The other significant neutron poison is $^{14}$N produced by $^{13}\mathrm{C}(\mathrm{p},\gamma)^{14}$N
following proton ingestion.
The competition between neutron production and capture results in a high 
plateau of $n_\mathrm{n}\sim (3$--$5) \times 10^{14}\,\mathrm{cm}^{-3}$. This high plateau is maintained for 
$\sim 10^5\,$s until $^{13}$C is burned away. Subsequently $n_\mathrm{n}$ 
decreases until a low plateau at $n_\mathrm{n}\sim 10^{11}\,\mathrm{cm}^{-3}$ is reached on a time scale of
$\sim 10^6\,$s. This low plateau is established through competition 
between neutron production by $^{17}\mathrm{O}(\alpha,\mathrm{n})^{20}\mathrm{Ne}$
and capture by the primary $^{12}$C and $^{16}$O, and is maintained for 
$\sim 10^7\,$s. The final evolution of $n_\mathrm{n}$ mainly follows the rate of
$^{17}\mathrm{O}(\alpha,\mathrm{n})^{20}$Ne, which in turn follows the temperature of 
the zone with some significant drop near the end of the star's life due to expansion 
after core C depletion and an eventual sharp rise due to contraction 
in the last $\sim (2$--$3)\times 10^{5}\,$s before core collapse. 
We note that $^{22}$Ne is also produced by
$^{14}$N$(\alpha,\gamma)^{18}$F$(e^{+}\nu_e)^{18}$O$(\alpha,\gamma)^{22}$Ne in the He shell due to proton ingestion. Neutron production
by $^{22}\mathrm{Ne}(\alpha,\mathrm{n})^{25}$Mg, however, is negligible compared to that by $^{13}$C$(\alpha,\mathrm{n})^{16}$O and
$^{17}\mathrm{O}(\alpha,\mathrm{n})^{20}\mathrm{Ne}$.

\begin{figure}[h]
\centerline{\includegraphics[width=85mm]{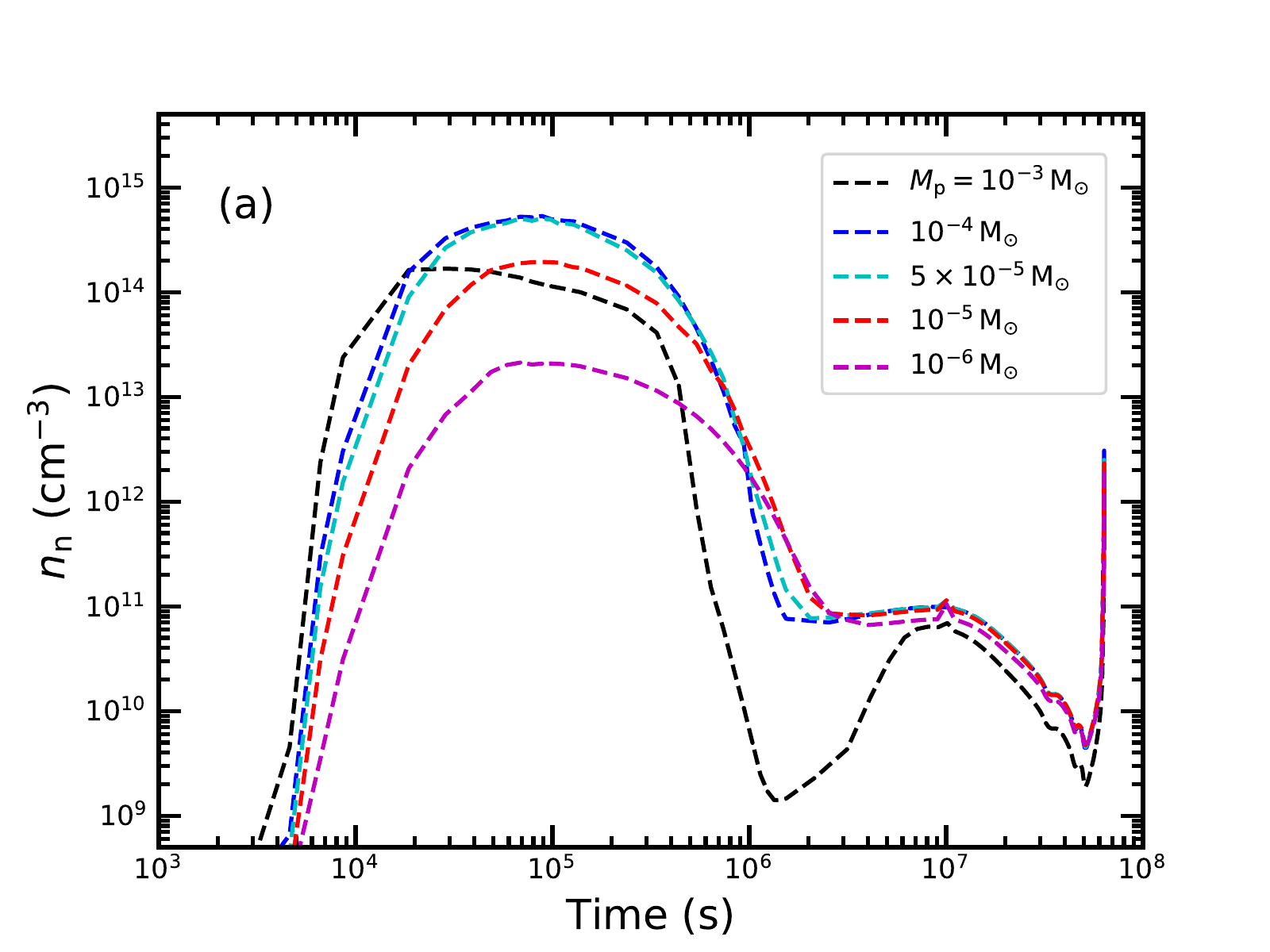}}
\centerline{\includegraphics[width=85mm]{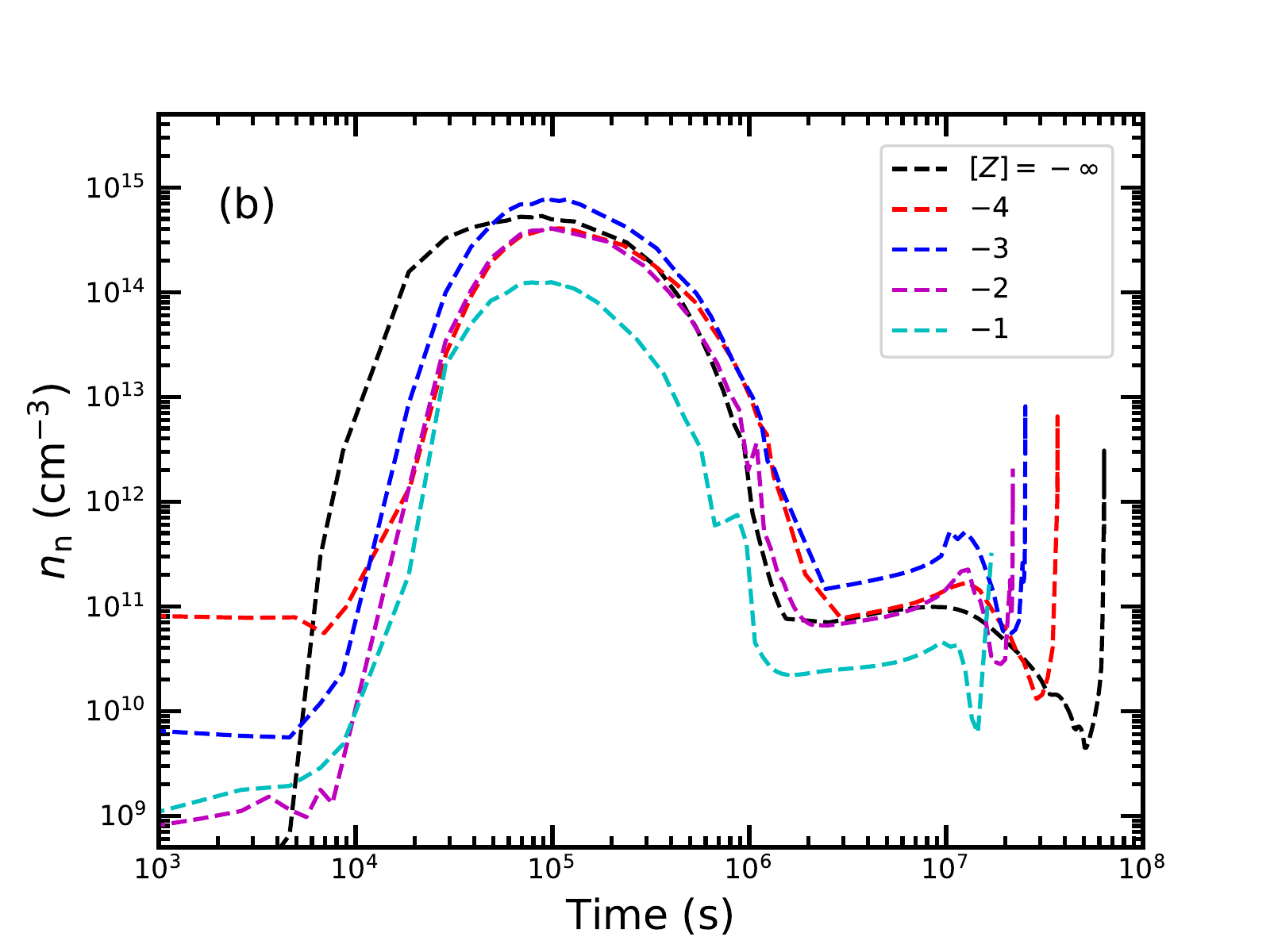}}
\caption{Neutron density $n_\mathrm{n}$ in a single zone at the base of the He shell
of a $25\,\mathrm{M}_\odot$ star with metallicity $[Z]$ as a function of time after a total mass $M_\mathrm{p}$ of
protons are ingested at central C depletion. (a) Cases of $M_\mathrm{p}=10^{-3}$ 
(\emph{black}), $10^{-4}$ (\emph{blue}), $5\times 10^{-5}$ (\emph{cyan}), $10^{-5}$ (\emph{red}), and 
$10^{-6} \,\mathrm{M}_\odot$ (\emph{magenta}), all for zero metallicity.
(b) Cases of zero metallicity (\emph{black}), $[Z]=-4$ (\emph{red}), $-3$ (\emph{blue}), 
$-2$ (\emph{magenta}), and $-1$ (\emph{cyan}), all for $M_\mathrm{p}=10^{-4} \,\mathrm{M}_\odot$.}
\label{fig:neut1}
\end{figure}

\begin{figure}
\centerline{\includegraphics[width=85mm]{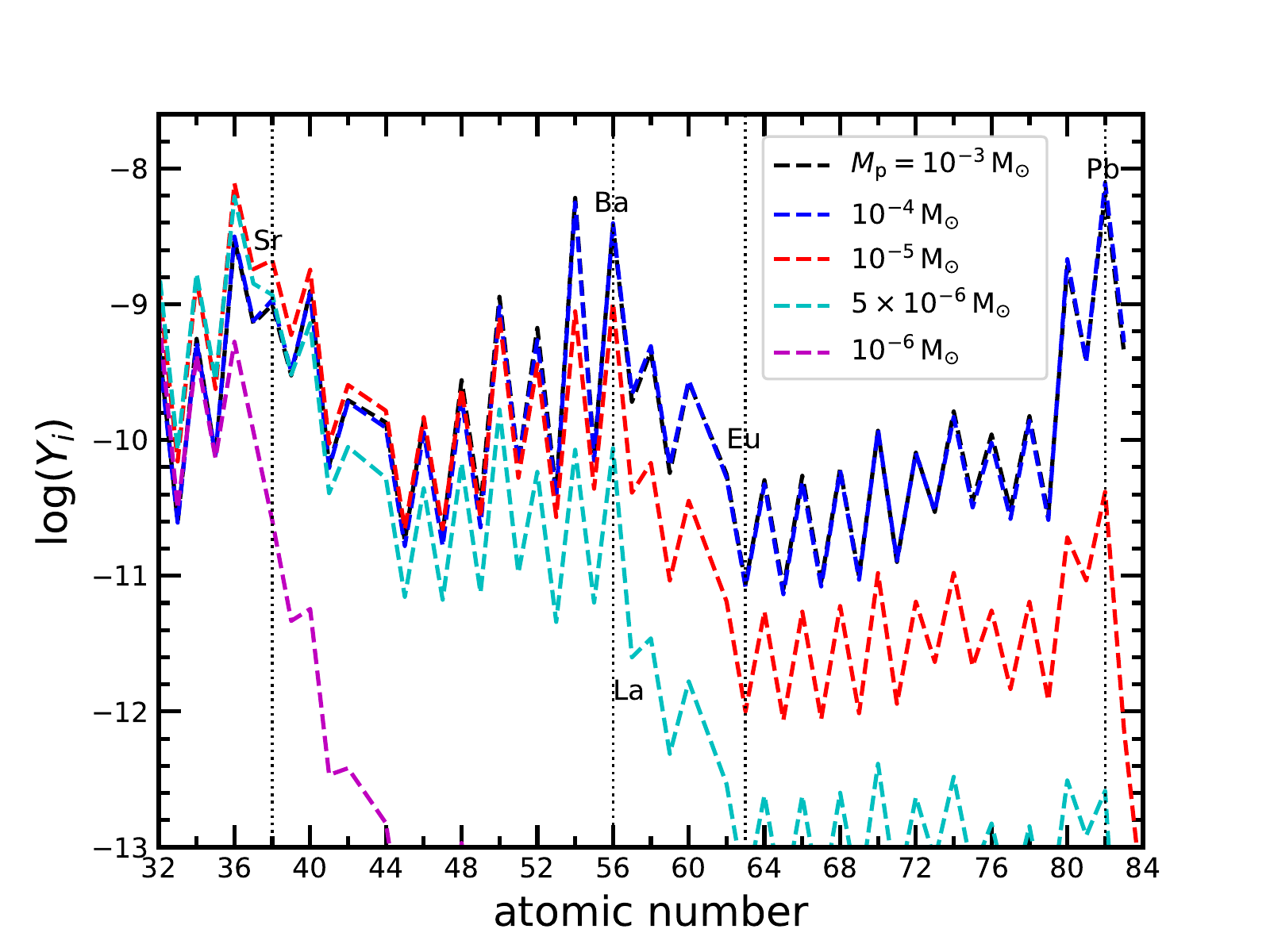}}
\caption{Yield patterns for a $25\,\mathrm{M}_\odot$ star with $[Z]=-3$ and ingestion of 
$10^{-3} \,\mathrm{M}_\odot$ (\emph{black}), $10^{-4} \,\mathrm{M}_\odot$ (\emph{blue}),
$10^{-5} \,\mathrm{M}_\odot$ (\emph{red}), $5\times10^{-6} \,\mathrm{M}_\odot$ (\emph{cyan}), and $10^{-6} \,\mathrm{M}_\odot$ (\emph{magenta}) 
of protons at central C depletion. Note that the black and blue curves are indistinguishable.}
\label{fig:v25pyield}
\end{figure}

Figure~\ref{fig:neut1}a compares the detailed evolution of $n_\mathrm{n}$ in the above zone when 
different amounts of protons are ingested in a $25\,\mathrm{M}_\odot$ star with zero metallicity.
For an ingested proton mass of $M_\mathrm{p}\lesssim 5\times 10^{-5} \,\mathrm{M}_\odot$, the high plateau
of $n_\mathrm{n}$ decreases monotonically with $M_\mathrm{p}$. For $M_\mathrm{p}\lesssim 10^{-6} \,\mathrm{M}_\odot$, 
however, the peak $n_\mathrm{n}$ is so low that nucleosynthesis by neutron capture 
effectively ceases. 
On the other hand, for $M_\mathrm{p}> 10^{-4} \,\mathrm{M}_\odot$, abundant protons start to produce
significant amounts of $^{14}$N through $^{13}\mathrm{C}(\mathrm{p},\gamma)^{14}\mathrm{N}$, which
then drives down $n_\mathrm{n}$ through $^{14}\mathrm{N}(\mathrm{n},\mathrm{p})^{14}$C. The effect of increased $^{14}\mathrm{N}$ is somewhat offset 
by higher $^{13}\mathrm{C}$ production such that neutron capture is still efficient for $M_\mathrm{p}\sim 10^{-3}$.
Overall, we find that for $10^{-5} \,\mathrm{M}_\odot\lesssim M_\mathrm{p}\lesssim 10^{-3} \,\mathrm{M}_\odot$,
neutron capture is efficient with total neutron exposures of 
$\int n_\mathrm{n}v_\mathrm{n}\,\mathrm{d}t\sim 15$--$45\,(\mathrm{mb})^{-1}$, where $v_\mathrm{n}$ is the thermal velocity of neutrons. 
This neutron capture results in production of heavy elements up to Bi with $[\textrm{Sr}/\textrm{Ba}]$ $\lesssim -0.5$
and with a comparable and sometimes higher amount of Pb relative to Ba. 
For $10^{-6} \,\mathrm{M}_\odot\lesssim M_\mathrm{p}<10^{-5} \,\mathrm{M}_\odot$, 
neutron capture is much less efficient but can still produce a reasonable amount of 
Sr with low Ba ($[\mathrm{Sr}/\mathrm{Ba}] > -0.5$) and negligible Pb 
(see Fig.~\ref{fig:v25pyield} and Table~\ref{tab:1}).

The metallicity of the star has little effect on 
the neutron density $n_\mathrm{n}$ for $[Z]\lesssim-2$ for a fixed $M_\mathrm{p}$ (see Fig.~\ref{fig:neut1}b). 
This is because the 
main neutron poison is the primary $^{16}$O from He burning in He shells of
such low metallicities. 
As $[Z]$ increases beyond $-2$, however, the peak $n_\mathrm{n}$
and the duration of the high plateau start to decrease due to the increasing 
amount of poisons in the star's birth material. This results in decreasing efficiency 
of nucleosynthesis by neutron capture with increasing metallicity. Nevertheless, 
such nucleosynthesis is still efficient up to $[Z]\sim -1$, 
producing mainly Sr along with elements up to around Ba.
The corresponding yields for $M_\mathrm{p}\gtrsim 5\times 10^{-5}  \,\mathrm{M}_\odot$ can be higher than  
those for the well-known weak \textsl{s}-process \citep{kappeler} at such metallicities 
(see \S\ref{sec:Z1} and Table~\ref{tab:1}).

We note that the values of $n_\mathrm{n}$ achieved 
during the high plateau correspond to typical 
\textsl{i}-process neutron densities, which have also been found in one-zone 
calculations of nucleosynthesis from proton 
ingestion in low to intermediate-mass stars \citep{dardelet2014} and in multi-zone calculations
for similar environments \citep{herwig2011,deni2017}. Whereas such $n_n$ values are determined ultimately by
the relevant nuclear physics and astrophysical conditions, it is important to understand how similar conditions
can be achieved in different environments, which may have very distinct implications for chemical evolution.
On the other hand, the values of $n_\mathrm{n}$ achieved during the low plateau is consistent 
with \textsl{s}-process neutron densities. Incidentally, $^{17}\mathrm{O}(\alpha,\mathrm{n})^{20}$Ne, which is responsible 
for maintaining the neutron density during this phase, is also known to be important for the weak
\textsl{s}-process in massive stars at low metallicites \citep{baraffe1992} 
and in spinstars \citep{pignatari2008,frischk2016}. Whereas neutrons are generated by 
$^{22}\mathrm{Ne}(\alpha, \mathrm{n})^{25}$Mg in the latter two cases, $^{17}$O$(\alpha,\mathrm{n})^{20}$Ne serves to
recover the neutron captured by $^{16}$O, thereby effectively reducing its neutron-capture capability by 
a factor of $\kappa \sim 13$--15. This effect is achieved because the rate of $^{17}\mathrm{O}(\alpha,\mathrm{n})^{20}$Ne
is higher than that of $^{17}\mathrm{O}(\alpha,\gamma)^{21}$Ne by a factor of $\kappa$ at temperatures of 
$\sim (2.5$--$3) \times 10^8\,$K typical of the weak \textsl{s}-process.

\subsection{Neutron Capture in Metal-Poor Massive Stars}
Based on the preceding discussion, metal-poor stars with $[Z]\lesssim -2$ but non-zero 
metallicities have similarly high plateaus of $n_\mathrm{n}$. Starting from
the Fe group nuclei in their birth material, nucleosynthesis by neutron capture 
readily produces heavy elements beyond Fe up to Bi.
As the initial metallicity increases from zero to $[Z]\sim -2$, more seeds become
available, and therefore, the amount of neutron-capture elements produced also
increases. For convenience of comparison with observational data, we use
$[\textrm{Ba}/\textrm{Eu}]$ as a proxy for the abundance pattern produced. For reference, contributions
to the solar abundances from the main \textsl{s}-process in AGB stars are 
characterized by 
$[\textrm{Ba}/\textrm{Eu}]_\mathrm{s}\sim 1.6$, whereas those from the \textsl{r}-process by $[\textrm{Ba}/\textrm{Eu}]_\mathrm{r}\sim -0.8$.

\begin{figure}
\centerline{\includegraphics[width=85mm]{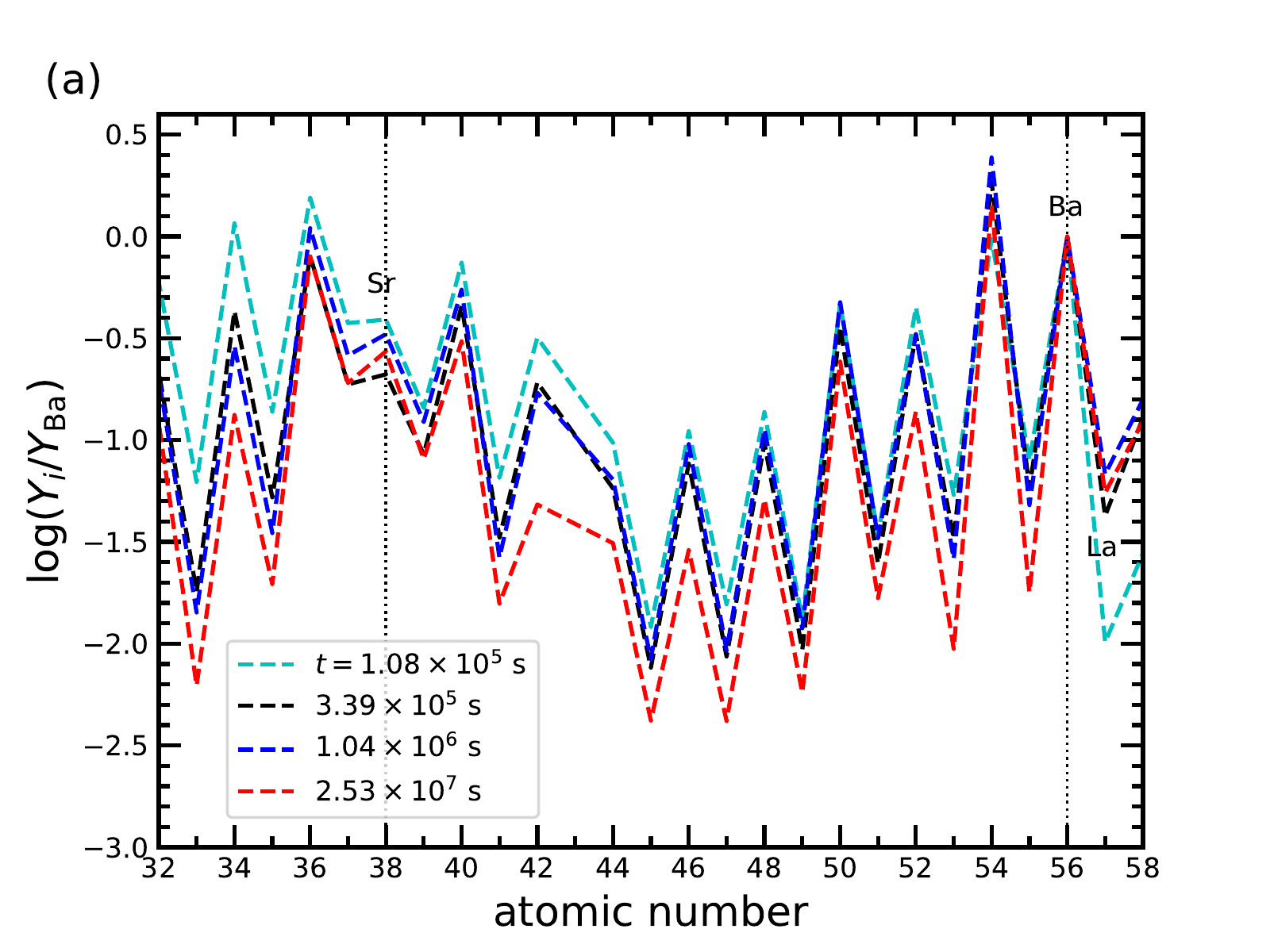}}
\centerline{\includegraphics[width=85mm]{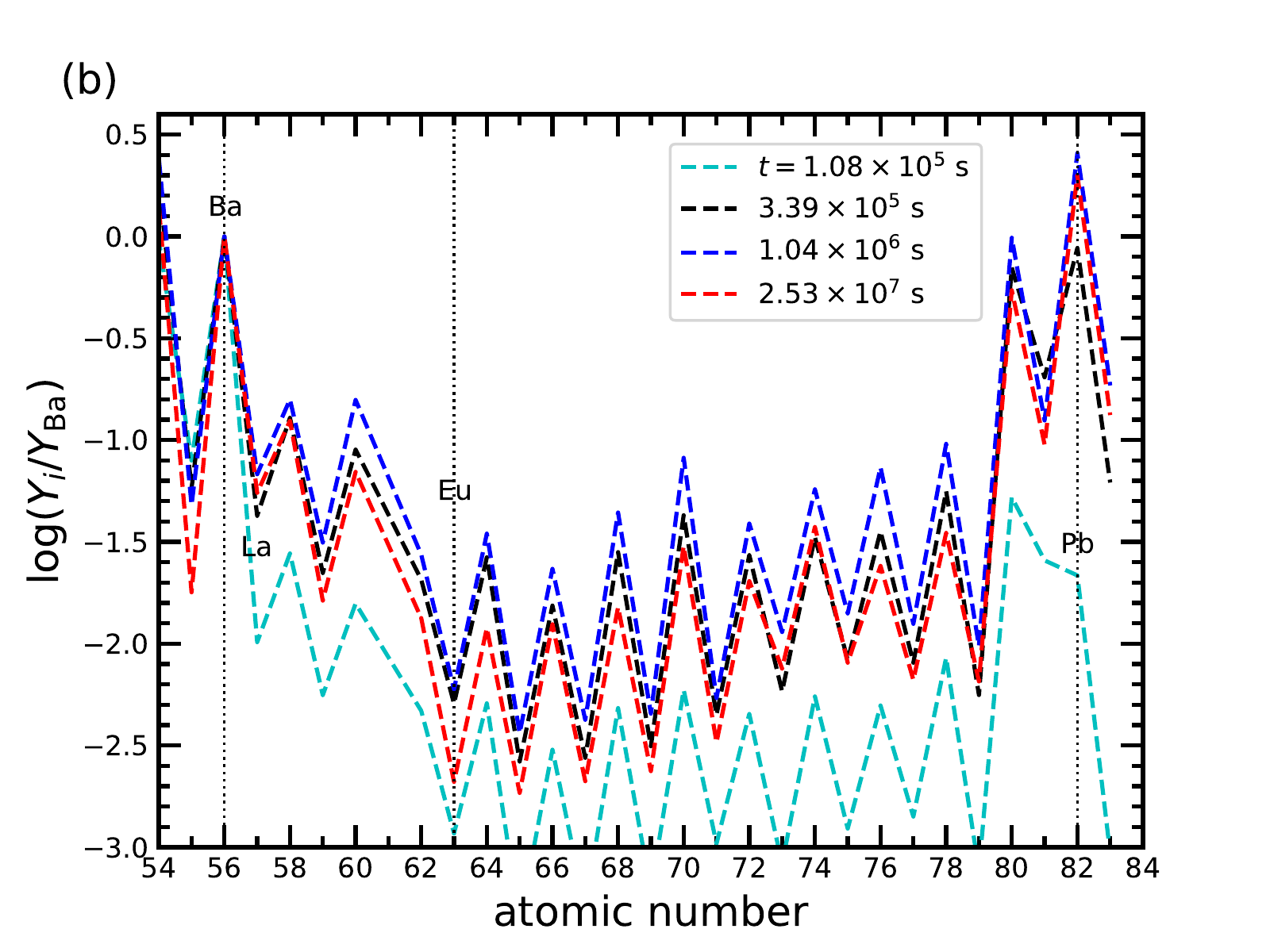}}
\caption{(a) Yield pattern from Ge to Ce normalized to Ba for a $25\,\mathrm{M}_\odot$ star with $[Z]=-3$ 
at time $t=1.08\times 10^{5}\,$s (\emph{cyan}), $3.39\times 10^{5}\,$s (\emph{black}), 
$1.04\times 10^{6}\,$s (\emph{blue}), and $2.53\times 10^{7}\,$s 
(\emph{red}, just before core collapse) after $10^{-4} \,\mathrm{M}_\odot$ of protons are ingested
at central C depletion. (b) Same as (a) but showing yield pattern from Xe to Bi.}
\label{fig:v25p2snap}
\end{figure}

Most of the nucleosynthesis is associated with the high plateau of $n_\mathrm{n}$
and occurs during the first $\sim 10^6\,$s after the proton ingestion. This stage 
contributes $\sim 98\,\%$ of the total neutron exposure.
For $10^{-5} \,\mathrm{M}_\odot\lesssim M_\mathrm{p}\lesssim 10^{-3} \,\mathrm{M}_\odot$,
the production pattern during this stage with $n_\mathrm{n}\sim (3$--$5) \times 10^{14}\,\mathrm{cm}^{-3}$
has the lowest $[\textrm{Ba}/\textrm{Eu}]$, as low as  $\sim 0.3$, which lies between $[\textrm{Ba}/\textrm{Eu}]_\mathrm{r}$ and $[\textrm{Ba}/\textrm{Eu}]_\mathrm{s}$.
When $n_\mathrm{n}$ drops to $\sim 10^{11}\,\mathrm{cm}^{-3}$ corresponding to the low plateau, however, 
neutron capture can still
continue for $\gtrsim 10^7\,$s.  During this time, abundance peaks at Sr, Ba, and Pb
corresponding to nuclei with magic neutron numbers $N=50$, $82$, and $126$ grow 
marginally due to very low neutron exposure, 
whereas the pattern between the peaks evolves considerably to resemble that of a main 
\textsl{s}-process. For example, $[\textrm{Ba}/\textrm{Eu}]$ can increase to $\sim 1.1$. 
The sharp rise in $n_\mathrm{n}$ during the last $\sim (2$--$3)\times10^5\,$s
does not contribute much to the total yield or the abundance peaks, but changes the 
abundances of those elements close to the peaks, such as La and Eu. As a result, 
$[\textrm{Ba}/\textrm{Eu}]$ can change to $\sim 0.9$ due to increase in the Eu abundance. 
Figure~\ref{fig:v25p2snap} shows the abundance patterns produced by neutron capture at 
different times after $10^{-4} \,\mathrm{M}_\odot$ of protons are ingested at central C depletion
in a $25 \,\mathrm{M}_\odot$ star with $[Z]=-3$.

\begin{figure}
\centerline{\includegraphics[width=85mm]{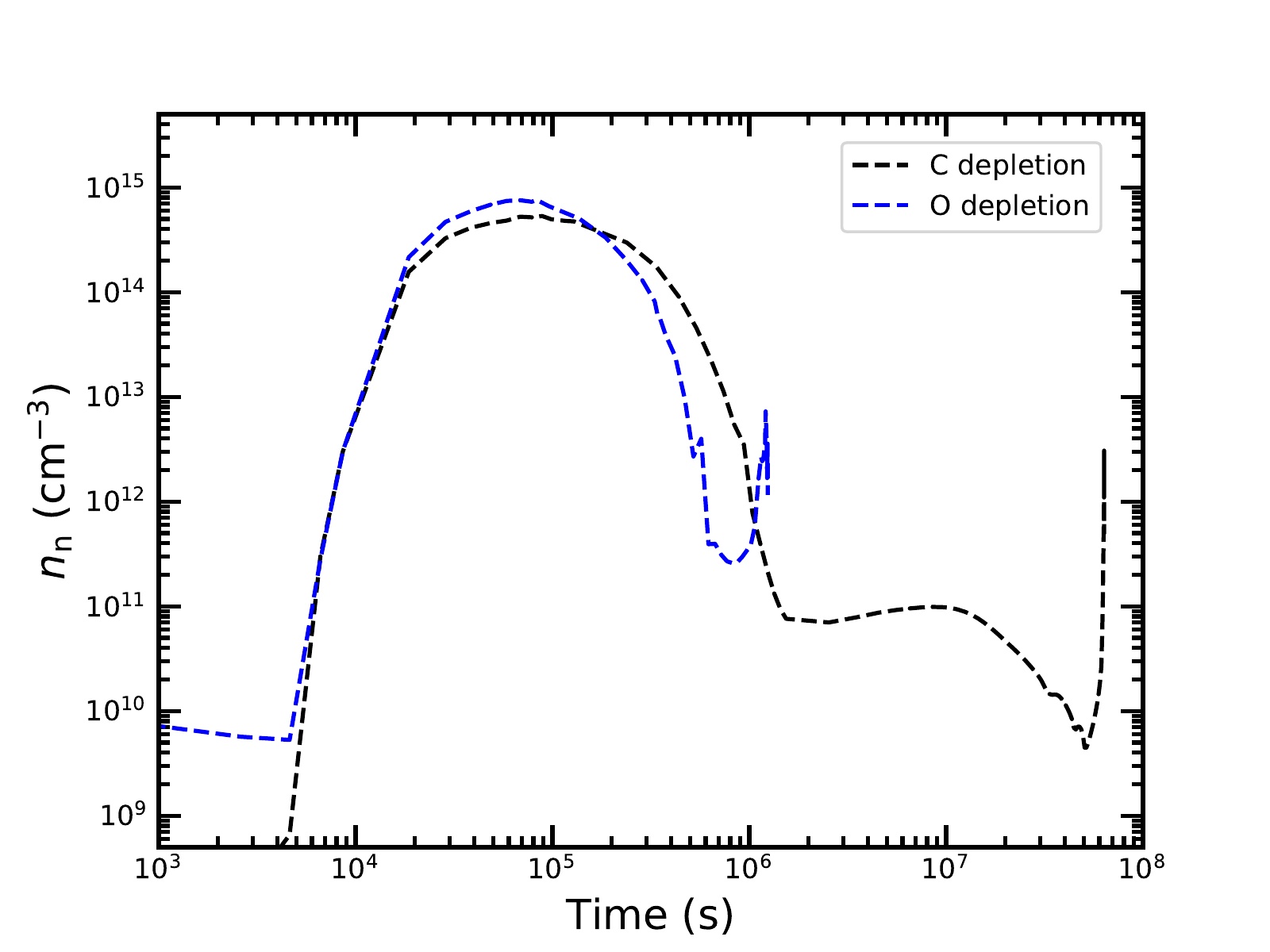}}
\caption{Neutron density $n_\mathrm{n}$ in a single zone at the base of the He shell 
of a metal-free $25\,\mathrm{M}_\odot$ star as a function of time after $10^{-4} \,\mathrm{M}_\odot$ of protons
are ingested at central O depletion (\emph{blue}) compared to that for ingestion at 
central C depletion (\emph{black}).}
\label{fig:z25codep}
\end{figure}

\begin{figure}
\centerline{\includegraphics[width=85mm]{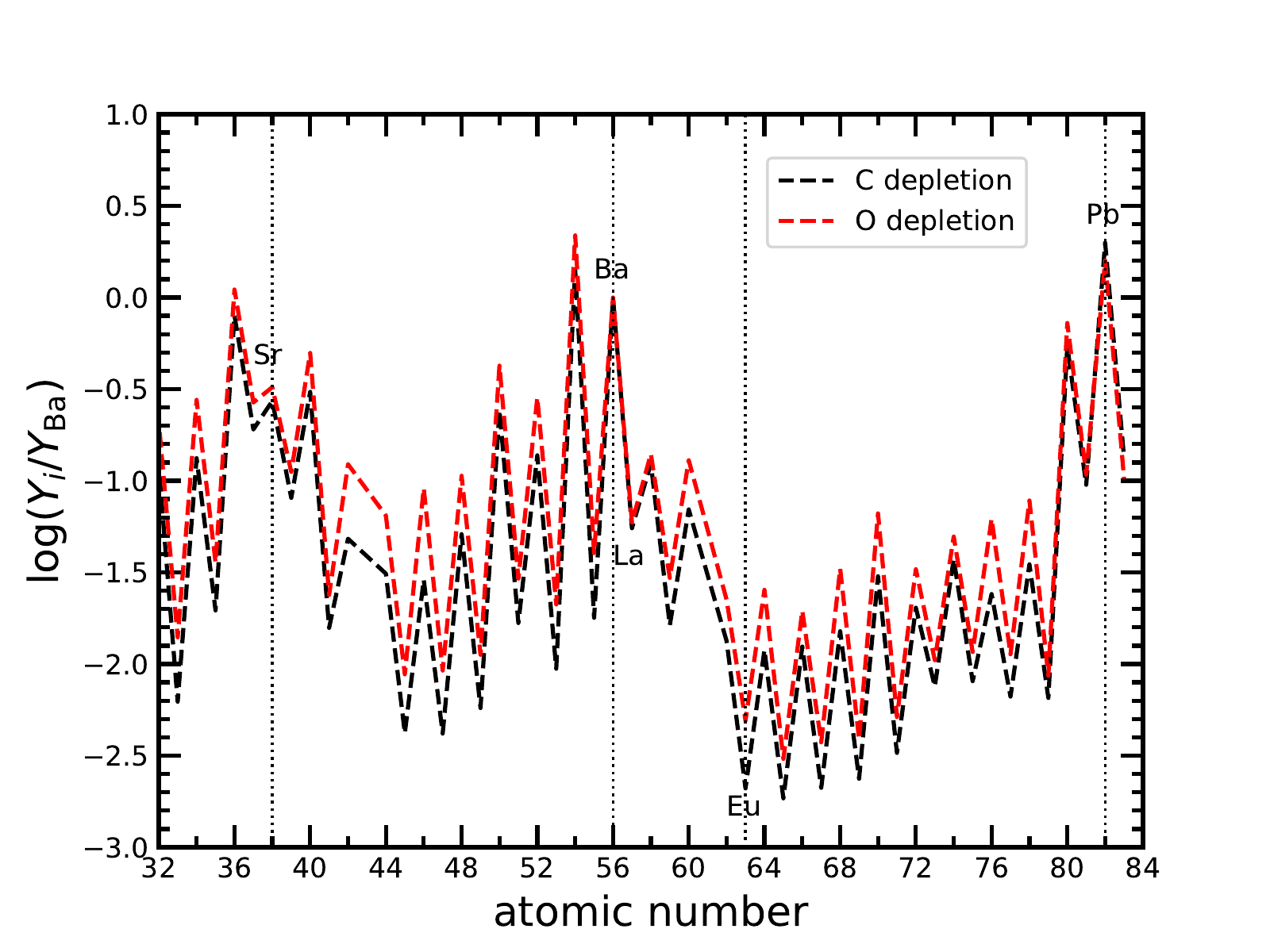}}
\caption{Yield pattern normalized to Ba for a $25\,\mathrm{M}_\odot$ star with $[Z]=-3$ and ingestion of 
$10^{-4} \,\mathrm{M}_\odot$ of protons at central C depletion (\emph{black}) compared to that for
ingestion at central O depletion (\emph{red}).}
\label{fig:v25codep}
\end{figure}

The above discussion assumes that protons are ingested when C is depleted in the 
center of the star. In order to explore nucleosynthesis from proton ingestion at a later
time, we ingest protons when O is depleted in the center and repeat the calculations. 
For our $25 \,\mathrm{M}_\odot$ models, the time of ingestion is $\sim 10^6\,$s prior to core collapse.
The evolution of $n_\mathrm{n}$ is similar to the case of proton ingestion at central C depletion,
with a high plateau at $n_\mathrm{n}\sim (3$--$5) \times 10^{14}\,\mathrm{cm}^{-3}$ lasting $\sim 10^5\,$s 
and a sharp rise
near the end of the star's life, except that there is not a low plateau but only a dip at 
$n_\mathrm{n}\sim 10^{11}\mathrm{cm}^{-3}$ prior to the sharp rise (see Fig.~\ref{fig:z25codep}).
Based on the discussion above, the corresponding nucleosynthesis should produce
similar relative abundance peaks but significantly different patterns between the
peaks compared to the case of proton ingestion at central C depletion. Specifically,
the lack of an extended low $n_\mathrm{n}$ plateau in the present case results in a lower $[\textrm{Ba}/\textrm{Eu}]$
of $\sim 0.6$ (see Fig.~\ref{fig:v25codep}).
As a limit for cases of later proton ingestion, $[\textrm{Ba}/\textrm{Eu}]$ could be as low as $\sim 0.3$ 
if nucleosynthesis occurs entirely during the phase of the high $n_\mathrm{n}$ plateau.

Therefore, the detailed abundance pattern produced by 
neutron capture in metal-poor massive stars 
is sensitive to the total duration $\Delta$ from the 
time of proton ingestion until core collapse. In general, 
$\Delta \lesssim 10^6\,$s produces $[\textrm{Ba}/\textrm{Eu}]\lesssim 0.6$
because the low neutron density phase is avoided. On the other hand, for $\Delta > 10^6\,$s, 
$[\textrm{Ba}/\textrm{Eu}]$ progressively increases towards the \textsl{s}-process value. 
This suggests that proton ingestion leading to $n_\mathrm{n}\gtrsim 10^{14}\,\mathrm{cm}^{-3}$ 
typical of the \textsl{i}-process may not necessarily produce $[\textrm{Ba}/\textrm{Eu}]$ values 
corresponding to the \textsl{r}/\textsl{s} stars. The initial phase of high neutron density is usually followed by a low neutron 
density phase, which can be avoided only if nucleosynthesis is terminated after the initial phase. In massive 
stars, this termination can happen naturally due to the onset of core collapse. In other scenarios of the
\textsl{i}-process such as associated with low and intermediate-mass stars, the mechanism for the termination would be 
very different. Regardless of the mechanism, the timescale for shutting down nucleosynthesis has to be 
$\ll 10^7\,$s in order to retain $[\mathrm{Ba}/\mathrm{Eu}] \lesssim 0.6$.

\begin{figure*}
\centerline{\includegraphics[width=\textwidth]{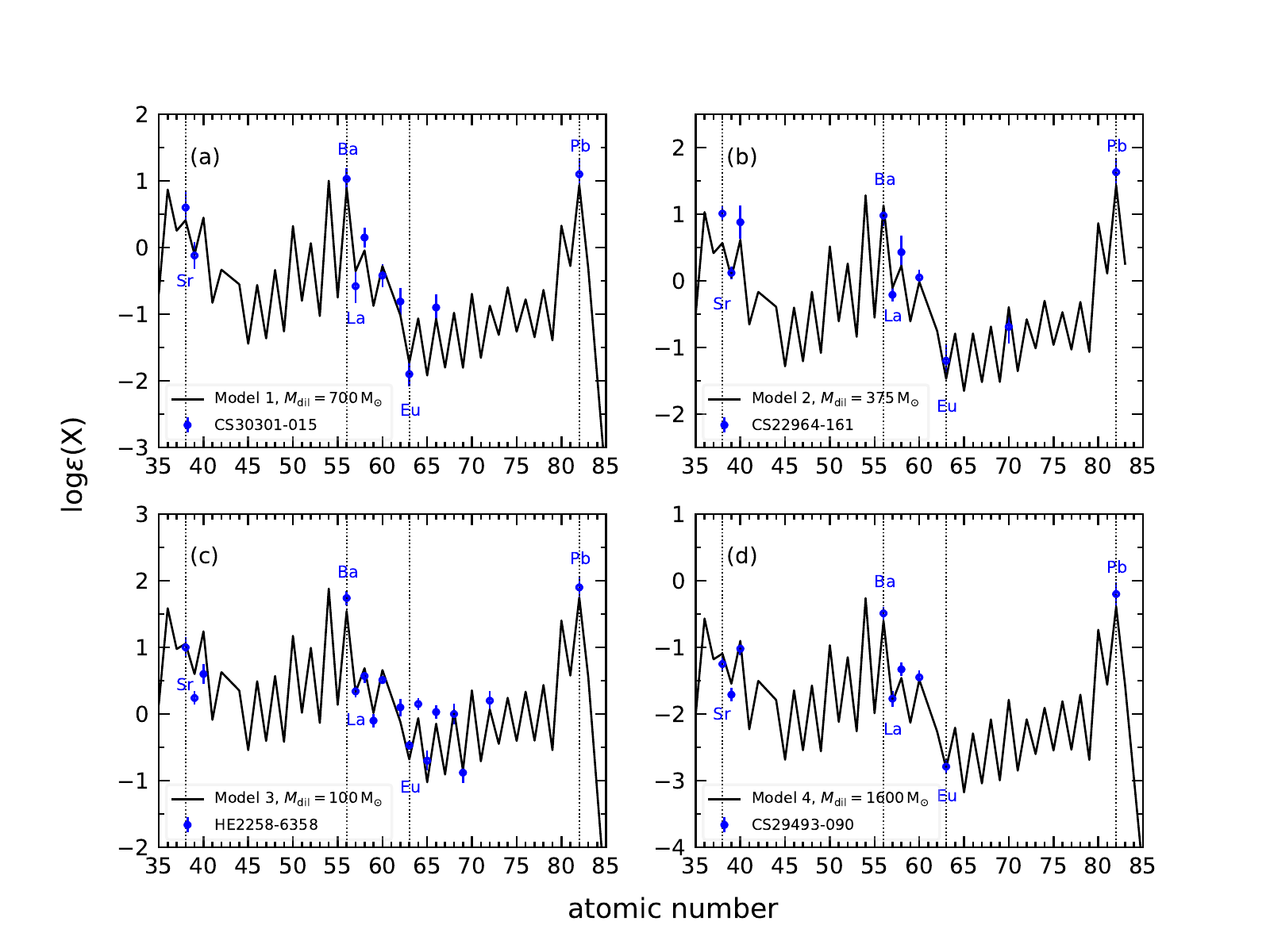}}
\caption{Comparison of model yields with abundances in VMP stars.
All models are for a $25\,\mathrm{M}_\odot$ star but differ in metallicity $[Z]$, ingested proton mass $M_\mathrm{p}$, 
ingestion time $t_{\rm ing}$, and explosion energy $E_{\rm expl}$.
Models 1, 2, 3, and 4 have $([Z]$, $M_\mathrm{p}/\,\mathrm{M}_\odot$, $t_{\rm ing}$, $E_{\rm expl}/{\rm ergs})=
(-3$, $5\times10^{-5}$, Cdep, $10^{50})$, 
$(-3$, $10^{-4}$, Cdep, $10^{50})$, 
$(-3$, $10^{-4}$, Odep, $10^{50})$, and 
$(-4$, $10^{-4}$, Odep, $3\times10^{50})$, respectively,
where Cdep (Odep) stands for central C (O) depletion.
Data are given as $\log\epsilon({\rm X})\equiv \log({\mathrm{X}/\mathrm{H}})+12$ for element X.
(a) Model 1 and CS 30301-015 \citep{aoki2002} for a dilution mass of $M_{\rm dil}=700 \,\mathrm{M}_\odot$.
(b) Model 2 and CS 22964-161 \citep{thompson2008} for $M_{\rm dil}=375 \,\mathrm{M}_\odot$.
(c) Model 3 and HE 2258-6358 \citep{placco2013} for $M_{\rm dil}=100 \,\mathrm{M}_\odot$.
(d) Model 4 and CS 29493-090 \citep{spite2014} for $M_{\rm dil}=1,\!600 \,\mathrm{M}_\odot$.}
\label{fig:obs}
\end{figure*}

\begin{figure*}
\centerline{\includegraphics[width=\textwidth]{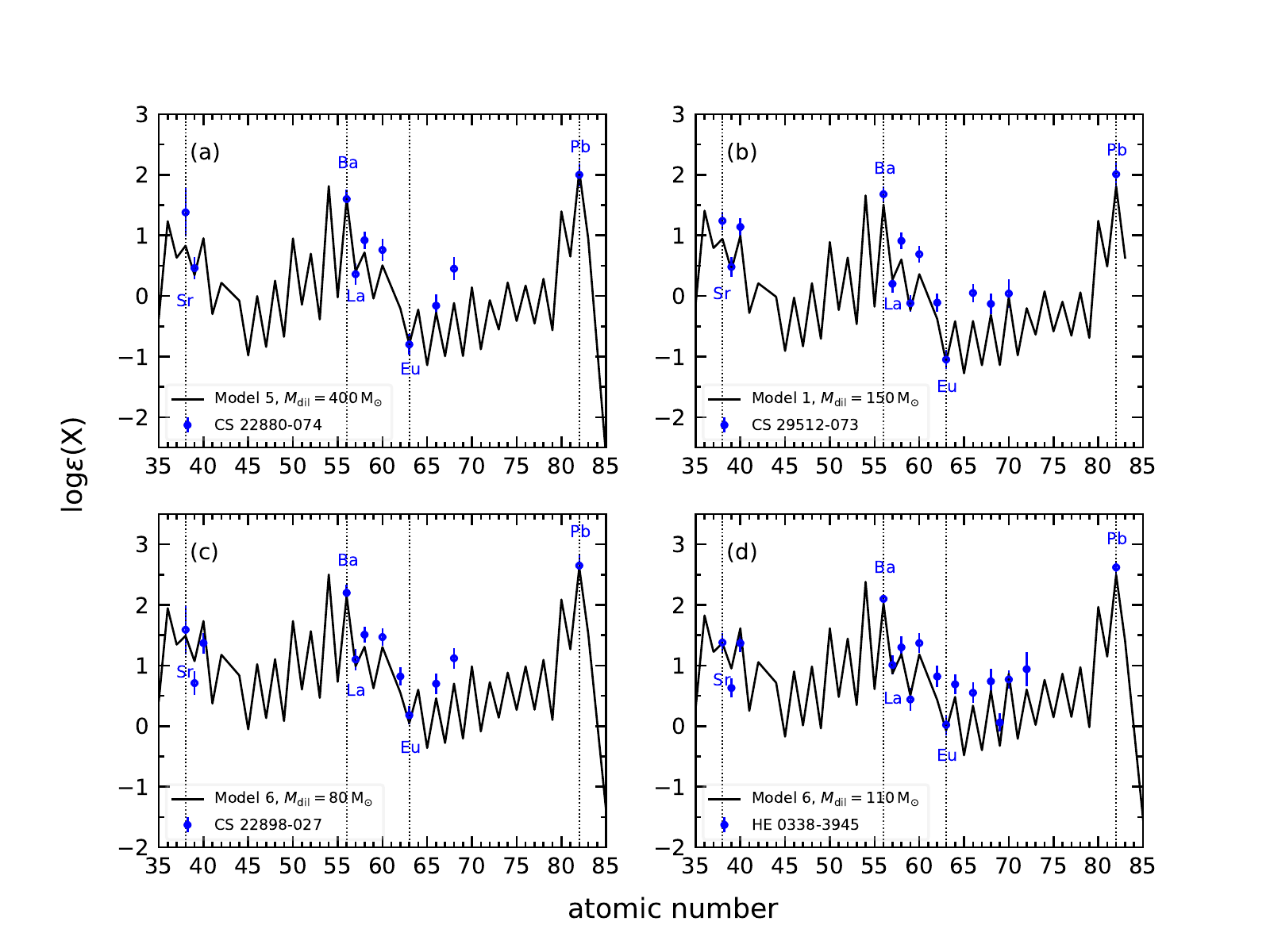}}
\caption{More comparison of model yields with abundances in VMP stars.
See Table~\ref{tab:2} for description of models.
(a) Model 5 and CS 22880-074 \citep{aoki2002} for a dilution mass of $M_{\rm dil}=400 \,\mathrm{M}_\odot$.
(b) Model 1 and CS 29512-073 \citep{roederer2014} for $M_{\rm dil}=150 \,\mathrm{M}_\odot$.
(c) Model 6 and CS 22898-027 \citep{aoki2002} for $M_{\rm dil}=80 \,\mathrm{M}_\odot$.
(d) Model 6 and HE 0338-3945 \citep{jonsell2006} for $M_{\rm dil}=110 \,\mathrm{M}_\odot$.}
\label{fig:obs2}
\end{figure*}

We calculate the nucleosynthesis for four models of a $25\,\mathrm{M}_\odot$ star 
(see Table~\ref{tab:2}) and compare
the results with data on four VMP stars in Fig.~\ref{fig:obs}. These models
differ in metallicity $[Z]$, ingested proton mass $M_\mathrm{p}$, ingestion time $t_{\rm ing}$, 
and CCSN explosion energy $E_{\rm expl}$.
Models 1, 2, 3, and 4 have $([Z],M_\mathrm{p}/\,\mathrm{M}_\odot$, $t_{\rm ing}$, $E_{\rm expl}/{\rm ergs})$ 
$=(-3$, $5\times10^{-5}$, $\mbox{Cdep}$, $10^{50})$, 
$(-3$, $10^{-4}$, Cdep, $10^{50})$, 
$(-3$, $10^{-4}$, Odep, $10^{50})$, and 
$(-4$, $10^{-4}$, Odep, $3\times10^{50})$, respectively,
where Cdep and Odep stand for central C ($\Delta\sim 3\times 10^7\,$s) and
O ($\Delta\sim 10^6\,$s) depletion, respectively.
The $[Z]$ is chosen to be consistent with the $[\textrm{Fe}/\textrm{H}]$ of the comparison star, and the 
$E_{\rm expl}$ to provide the plausible dilution mass for
accommodating the observed absolute abundances (see \S\ref{sec-discuss}). 
For the chosen $E_{\rm expl}$, the CCSN explosion has little
effect on the yields of neutron-capture elements but can cause a minor decrease in $[\textrm{Ba}/\textrm{Eu}]$ by $\sim 0.1$ 
(see \S\ref{sec:shock} and Tables~\ref{tab:1}--\ref{tab:2}).
Figure~\ref{fig:obs} shows that the patterns from Ba to Pb produced by
the models are in excellent agreement with the data on the four VMP stars. 
Similar agreement is also found for other VMP stars (see Fig.~\ref{fig:obs2}).

\begin{figure}
\centerline{\includegraphics[width=85mm]{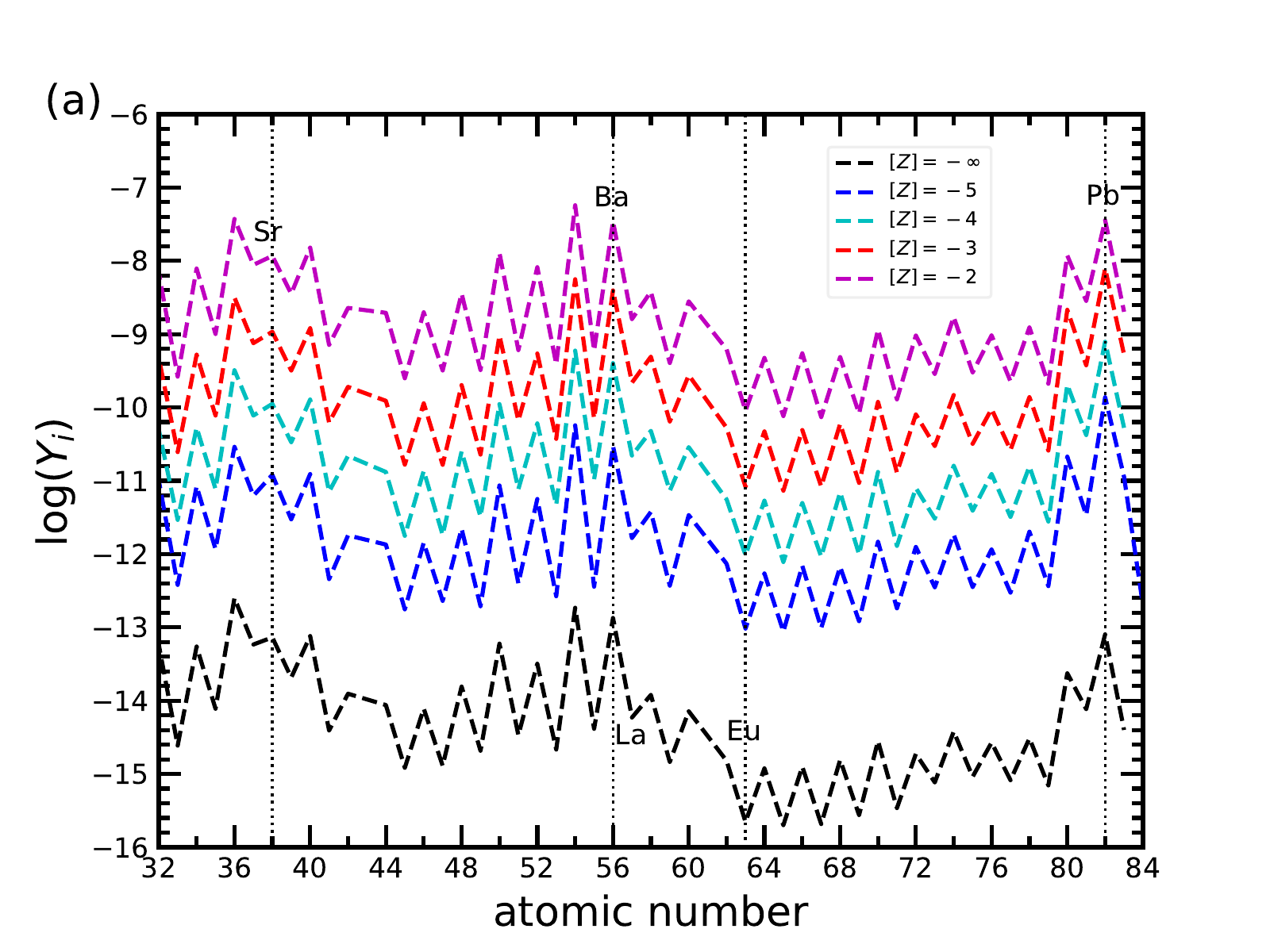}}
\centerline{\includegraphics[width=85mm]{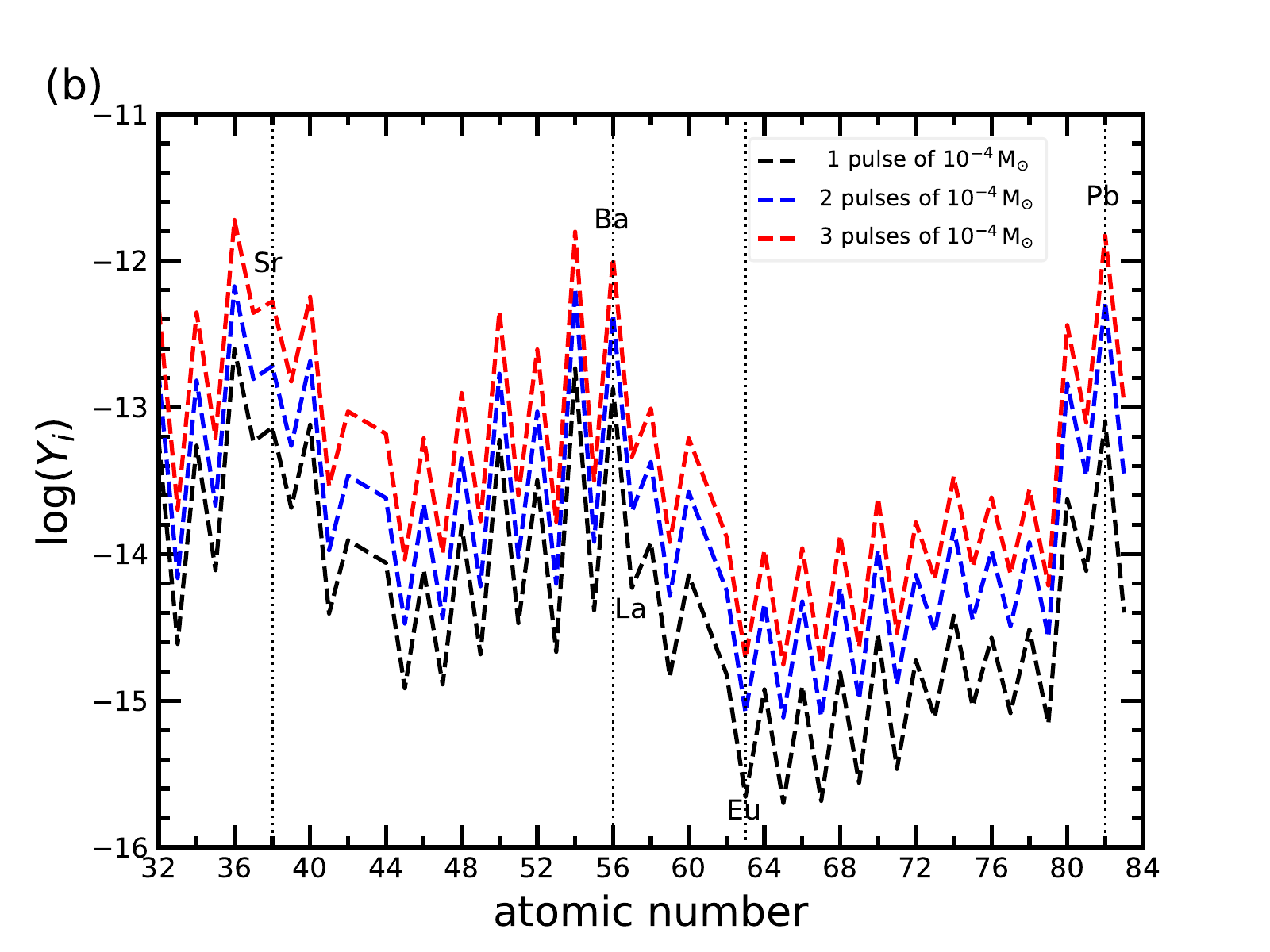}}
\caption{(a) Yield patterns for a $25\,\mathrm{M}_\odot$ star with ingestion of 
$10^{-4} \,\mathrm{M}_\odot$ of protons at central C depletion for
zero metallicity (\emph{black}), $[Z]=-5$ (\emph{blue}),
$-4$ (\emph{cyan}), $-3$ (\emph{red}), and $-2$ (\emph{magenta}).
(b) Yield pattern for the zero-metallicity case (\emph{black}) in (a)
compared to those for 2 (blue) and 3 (red) pulses of ingestion with 
$M_\mathrm{p}=10^{-4} \,\mathrm{M}_\odot$ for each pulse and an interval of $10^6\,$s
between successive pulses starting at central C depletion.}
\label{fig:p2Z}
\end{figure}

As mentioned above, because $n_\mathrm{n}$ is nearly independent of metallicity for $[Z]\lesssim -2$,
the total yield of heavy elements is proportional to the amount of seeds inherited 
from the initial metallicity (see Fig.~\ref{fig:p2Z}a). 
This relation between the yield and metallicity, however, breaks down for primordial massive
stars with zero metallicity, which we discuss below.

\subsection{Neutron Capture in Metal-Free Massive Stars}

Because the birth material of primordial stars is free of metals that can serve as seeds 
for neutron capture, no significant heavy element 
synthesis is expected even though $n_\mathrm{n}$ values are almost identical to those in
their metal-poor counterpart with $[Z] \lesssim -2$. Interestingly, however, 
tiny amounts of Ca isotopes 
(with a mass fraction of $\sim 10^{-10}$ for $^{40}$Ca and $\sim 10^{-11}$ for $^{44}$Ca)
and even less amounts of Ti isotopes (with mass fractions of $\sim 10^{-12}$ for
$^{46}$Ti to $^{50}$Ti) produced by the hot CNO cycle \citep{keller} are present in the He 
shell, and these primary nuclei act as the seeds. 
Because most of the seeds ($^{40}$Ca) have the magic neutron number $N=20$,
production of heavy elements beyond Fe is greatly slowed down due to the bottleneck
at the next magic number $N=28$. We find that even with the $n_\mathrm{n}$ of the high plateau, 
clearing this bottleneck takes most of the duration ($\sim 10^5\,$s) 
of this plateau, which limits the production of heavy elements. 
Nevertheless, elements up to Bi can be produced, 
with their yields particularly sensitive to
the peak $n_\mathrm{n}$, and hence the ingested proton mass $M_\mathrm{p}$ (see Fig.~\ref{fig:neut1}a).
The maximum yields occur for $M_\mathrm{p} \sim 10^{-4} \,\mathrm{M}_\odot$ but are
$\sim 500$ times lower than the yields of the $[Z]=-5$ counterpart or comparable to
those of the $[Z]\sim -7.5$ counterpart for the same $M_\mathrm{p}$. Nonetheless, 
metal-free stars with $M_\mathrm{p} \sim 10^{-4} \,\mathrm{M}_\odot$ can provide the lowest amounts of 
neutron-capture elements measured in VMP stars (see \S\ref{sec-discuss}).

In sharp contrast to metal-poor massive stars, metal-free ones leave
most of the seeds unused even for $M_\mathrm{p} \sim 10^{-4} \,\mathrm{M}_\odot$ with the highest 
efficiency of neutron capture. In addition, fresh seeds are produced by neutron capture 
on the primary $^{22}$Ne, which has a mass fraction of $\sim 2 \times 10^{-7}$ in the He shell. 
Were more episodes of proton ingestion to occur in metal-free massive stars,
the unused and freshly-synthesized seeds could be utilized to increase the yields of heavy
elements dramatically (see \S\ref{sec-multi}). For example, three episodes 
with $M_\mathrm{p}=10^{-4}\,\mathrm{M}_\odot$ for each would increase the yields by $\sim 10$ times. 

As mentioned above, \citet{clarkson2017} found that an \textsl{i}-process could result from proton 
ingestion in the convective He shell in a zero-metallicity star of $45 \,\mathrm{M}_\odot$. In their study 
based on the MESA code, a substantial amount of proton ingestion occurs just after the end of 
core He burning when the convective He shell exchanges material with the H-rich layers. This 
large proton ingestion leads to large energy generation with 
$\epsilon_{{\rm nuc}}\tau_{{\rm conv}}\approx 0.26\, E_{{\rm int}} $. In their one-zone 
nucleosynthesis calculations, the initial proton mass fraction was taken to be $1\,\%$, which produced 
a peak $n_\mathrm{n}$ of $\approx 6\times 10^{13}\,\mathrm{cm}^{-3}$. This value is approximately an order of magnitude 
lower than the highest $n_\mathrm{n}$ found in the present study. The difference is simply due to the larger 
proton ingestion used in \citet{clarkson2017}. For example, $M_\mathrm{p} \lesssim 10^{-4} \,\mathrm{M}_\odot$ in our 
calculations corresponds to a proton mass fraction of $\lesssim 0.01\,\%$ for a single-zone calculation,
to be compared with $1\,\%$ adopted by \citet{clarkson2017}. A mass fraction of $1\,\%$ would correspond to 
$M_\mathrm{p}\sim 10^{-2} \,\mathrm{M}_\odot$ for our study. As discussed above, for $M_\mathrm{p}\gtrsim 10^{-3} \,\mathrm{M}_\odot$,
substantial amounts of $^{14}$N are produced to make $^{14}$N the dominant 
neutron poison, which lowers the neutron abundance through $^{14}\mathrm{N}(\mathrm{n},\mathrm{p})^{14}$C.

\subsection{Multiple Episodes of Proton Ingestion}
\label{sec-multi}

With the overshoot prescription described in \S\ref{sec-method}, we found 
that subsequent to the proton ingestion at central C depletion, 
additional ingestion occurred at central O and Si depletion, especially for stars with $[Z]\lesssim -4$.
Because each episode of proton ingestion with $10^{-5} \,\mathrm{M}_\odot\lesssim M_\mathrm{p} \lesssim 10^{-3} \,\mathrm{M}_\odot$ 
leads to a period of high neutron density, multiple episodes of ingestion can 
significantly increase the efficiency of neutron-capture nucleosynthesis. This is 
particularly relevant for metal-free stars, where a substantial 
amount of the seeds are left behind and additional seeds are produced by
neutron capture following a single episode of proton ingestion. 
For these stars, with each additional episode of proton ingestion, 
the yields of heavy elements increase substantially 
as the unused seeds from the previous episode are utilized.

We explored multiple episodes of proton ingestion for a metal-free star by injecting
two or three pulses of protons with $M_\mathrm{p}=10^{-4} \,\mathrm{M}_\odot$ for each pulse and an interval of $10^6\,$s 
between successive pulses starting at central C depletion. 
The yields of heavy elements increased by a factor of $\sim 4$ or 
10, respectively, compared to the case of a single pulse (see Table~\ref{tab:1} and Fig.~\ref{fig:p2Z}b).
For metal-poor stars, most of the seeds were used up after a 
single pulse with $M_\mathrm{p}\gtrsim 10^{-5} \,\mathrm{M}_\odot$. 
Additional pulses could increase the yield of Pb relative to Ba 
and Sr (see Models 5 and 6 in Table~\ref{tab:2}), 
thereby providing overall patterns in excellent agreement with data on 
some CEMP-\textsl{s} and CEMP-\textsl{r}/\textsl{s} stars that are enhanced in Pb (see Fig.~\ref{fig:obs2}).

\subsection{Effect of the CCSN Shock}
\label{sec:shock}
As the shock goes through the He shell, 
it raises the temperature of the material there so that a short burst of neutrons 
may be released via $(\alpha,\mathrm{n})$ reactions on $^{13}$C, $^{17}$O, and $^{22}$Ne, which have been produced 
earlier due to proton ingestion. The subsequent neutron capture can affect the final $[\textrm{Ba}/\textrm{Eu}]$. 
The amount of neutrons released depends on the explosion energy, 
which controls the peak temperature of the shocked
material and hence the rate of the relevant $(\alpha,\mathrm{n})$ reactions. 
For an explosion energy of $E_{\rm expl}=1.2\times 10^{51}\,$ergs,
we found that the yield of Eu increased by a factor of $\sim 2$ but 
that of Ba was virtually unchanged, resulting in a decrease of
$[\textrm{Ba}/\textrm{Eu}]$ by $\sim 0.3$ compared to the pre-shock value. 
For lower $E_{\rm expl}$ of $10^{50}$ and $3\times 10^{50}\,$ergs, 
release of neutrons by the shock was so inefficient 
that $[\textrm{Ba}/\textrm{Eu}]$ only increased by $\sim 0.05$ and $0.1$, respectively, after passage of the shock.

\subsection{Neutron Capture in a $25\,\mathrm{M}_\odot$ Star with $[Z]=-1$}
\label{sec:Z1}
In addition to models with initial metallicity up to $[Z]=-2$, 
we also calculated the nucleosynthesis in a $25\,\mathrm{M}_\odot$ star with $[Z]=-1$. 
In this case, the initial metal composition was scaled from the solar abundances  
for all the corresponding elements up to Bi. The effect of neutron poisons inherited from 
the initial metallicity is evident from the lower neutron density compared to 
the $[Z]\lesssim -2$ models (see Fig.~\ref{fig:neut1}b). 
The final yields of heavy elements in this case include the 
contributions from the initial metallicity and the weak \textsl{s}-process in addition to 
what is produced due to proton ingestion (see  Table~\ref{tab:1}). We found that neutron capture was most 
efficient for an ingested proton mass of $M_\mathrm{p}\sim 10^{-3} \,\mathrm{M}_\odot$, 
producing substantial amounts of Sr and Ba. Efficiency of neutron capture decreased 
rapidly for lower amounts of proton ingestion, with production of mostly Sr. Nevertheless,
for $M_\mathrm{p}\gtrsim 5\times 10^{-5}\,\mathrm{M}_\odot$, the production of Sr due to proton ingestion still exceeds 
that by the weak \textsl{s}-process (see Table~\ref{tab:1}). 
Therefore, nucleosynthesis from proton ingestion in massive stars
is expected to have an important effect on Galactic 
chemical evolution of heavy elements at least up to $[\textrm{Fe}/\textrm{H}]\sim -1$.

\section{Implications for Observations of VMP Stars}
\label{sec-discuss}

Based on exploration with 1D models, we have proposed a new site for neutron-capture
nucleosynthesis in early stars of $\sim 20$--$30\,\mathrm{M}_\odot$. This nucleosynthesis results 
from proton ingestion in convective He shells and has characteristics  
ranging from those of the \textsl{i}-process, which has been
studied in connection with proton ingestion in stars of $<10\,\mathrm{M}_\odot$, to those of 
the main \textsl{s}-process in low and intermediate-mass stars as well as those of the weak 
\textsl{s}-process in massive stars. Our calculations suggest that the proposed
neutron-capture site can resolve a number of observational 
puzzles pertaining to VMP stars. Before we discuss these implications in detail,
we summarize the features of this site that are
important in comparison with observations. 

The amount of heavy elements produced from proton ingestion
in an early massive star depends mainly on the ingested proton mass $M_\mathrm{p}$ and 
the amount of seeds inherited from its initial metallicity $[Z]\sim[{\rm Fe/H}]$.
For $[\textrm{Fe}/\textrm{H}]\lesssim -2$, the yield scales linearly with the number ratio (Fe/H) except for 
metal-free stars, where the seeds are the primary Ca and Ti isotopes produced during their
evolution. For stars with a specific (including zero) metallicity, neutron capture is efficient for 
$10^{-5} \,\mathrm{M}_\odot\lesssim M_\mathrm{p}\lesssim 10^{-3} \,\mathrm{M}_\odot$ and is generally characterized by 
high Ba and Pb production with $[\textrm{Sr}/\textrm{Ba}]\lesssim -0.5$. In contrast, for 
$10^{-6} \,\mathrm{M}_\odot\lesssim M_\mathrm{p} <10^{-5} \,\mathrm{M}_\odot$, 
neutron capture is less efficient with low Ba and Pb production and $[\textrm{Sr}/\textrm{Ba}]> -0.5$. The yield 
patterns for $10^{-5} \,\mathrm{M}_\odot\lesssim M_\mathrm{p}\lesssim 10^{-3} \,\mathrm{M}_\odot$ can vary from 
\textsl{s}-like ($[\textrm{Ba}/\textrm{Eu}] > 0.6$ for $\Delta > 10^6\,$s) to \textrm{r}/\textrm{s}-like 
($[\textrm{Ba}/\textrm{Eu}] \lesssim 0.6$ for $\Delta \lesssim 10^6\,$s) depending on the time $\Delta$ available 
for neutron capture, with an overall range of $[\textrm{Ba}/\textrm{Eu}]\sim 0.2$--$1$.

\subsection{Ubiquity of {\rm Sr} and {\rm Ba}}

The ubiquity of Sr in VMP stars may not be surprising.
Neutrino-driven winds from a protoneutron star created in a CCSN can produce 
heavy elements up to $A\sim 120$ with $\sim10^{-8}$--$10^{-7}\,\mathrm{M}_\odot$ of Sr \citep{arcones2011}. 
When mixed with $\sim 10^2$--$10^4\,\mathrm{M}_\odot$ of ISM as expected for a wide range of 
CCSN explosion energy, such yields can explain 
$\log \epsilon({\rm Sr}) \equiv \log({\mathrm{Sr}/\mathrm{H}})+12\lesssim 0$ 
observed in extremely metal-poor stars with $[\textrm{Fe}/\textrm{H}]\lesssim -3.5$ \citep{saga}.
Ba is also frequently observed in such stars including several CEMP-no stars.
The latter belong to a group with some members considered 
to have formed from an ISM polluted exclusively by  primordial 
(metal-free) massive stars \citep{hansenCEMPno}. This presents
a major puzzle because Ba is not produced in neutrino-driven winds and nor is its widespread
presence accountable by the 
\textsl{r}-process in such rare events as neutron star mergers or jet-driven CCSNe. 
Our proposed site, however, operates in a significant fraction of massive stars
even at zero metallicity, so it can provide Ba 
to stars of the lowest metallicities including CEMP-no stars. 
Stars with $[\textrm{Fe}/\textrm{H}]\lesssim -3.5$ are observed to have
$\log \epsilon({\rm Ba}) \lesssim -1$ with the lowest measured
value of $\sim -3.9$. Our metal-free $25\,\mathrm{M}_\odot$ models with $M_\mathrm{p}\sim 10^{-4} \,\mathrm{M}_\odot$
can provide $\log \epsilon({\rm Ba})\sim -5$ to $-3$ if their Ba yields (see Table~\ref{tab:1}) 
are diluted by $\sim 10^2$--$10^4\,\mathrm{M}_\odot$ of ISM. With few additional episodes of 
proton ingestion, such models can provide
$\log \epsilon({\rm Ba})$ up to $\sim -2$, the maximum Ba
enrichment observed in stars with $[\textrm{Fe}/\textrm{H}]\lesssim -4$.
Extremely metal-poor massive stars with 
$-7\lesssim[\textrm{Fe}/\textrm{H}]\lesssim -4$ can account for
$-3<\log \epsilon({\rm Ba})\lesssim 0$, which includes the highest
Ba enrichment observed in stars with $[\textrm{Fe}/\textrm{H}]\lesssim -3.5$. 
Therefore, the proposed site in metal-free and 
metal-poor massive stars can explain adequately the ubiquity of Ba in VMP stars.

\subsection{Diversity of VMP Stars}

We note that the amount of Sr synthesized  
by neutrino-driven winds is more than that of Sr or Ba by the proposed site for massive stars 
with $[\textrm{Fe}/\textrm{H}]\lesssim -5$ and comparable to that for $[\textrm{Fe}/\textrm{H}]\sim -4$  (see Table~\ref{tab:1}).
The Sr from these winds along with the Ba from the proposed site at such
metallicities would naturally explain VMP stars with $[\textrm{Sr}/\textrm{Ba}]\gtrsim -0.5$ 
\citep{spite2014,saga}. 
Were the wind material not ejected such as in a CCSN with 
weak explosion and hence severe fallback, both the Sr and Ba in the ejecta would be 
exclusively from the proposed site operating in the He shell, which is
expected to be ejected even for weak explosion. In this case, both $[\textrm{Sr}/\textrm{Ba}]> -0.5$
and $[\textrm{Sr}/\textrm{Ba}]\lesssim -0.5$ can be produced depending on the ingested proton mass,
with the latter values consistent with other observations.

\begin{figure*}
\centerline{\includegraphics[width=\textwidth]{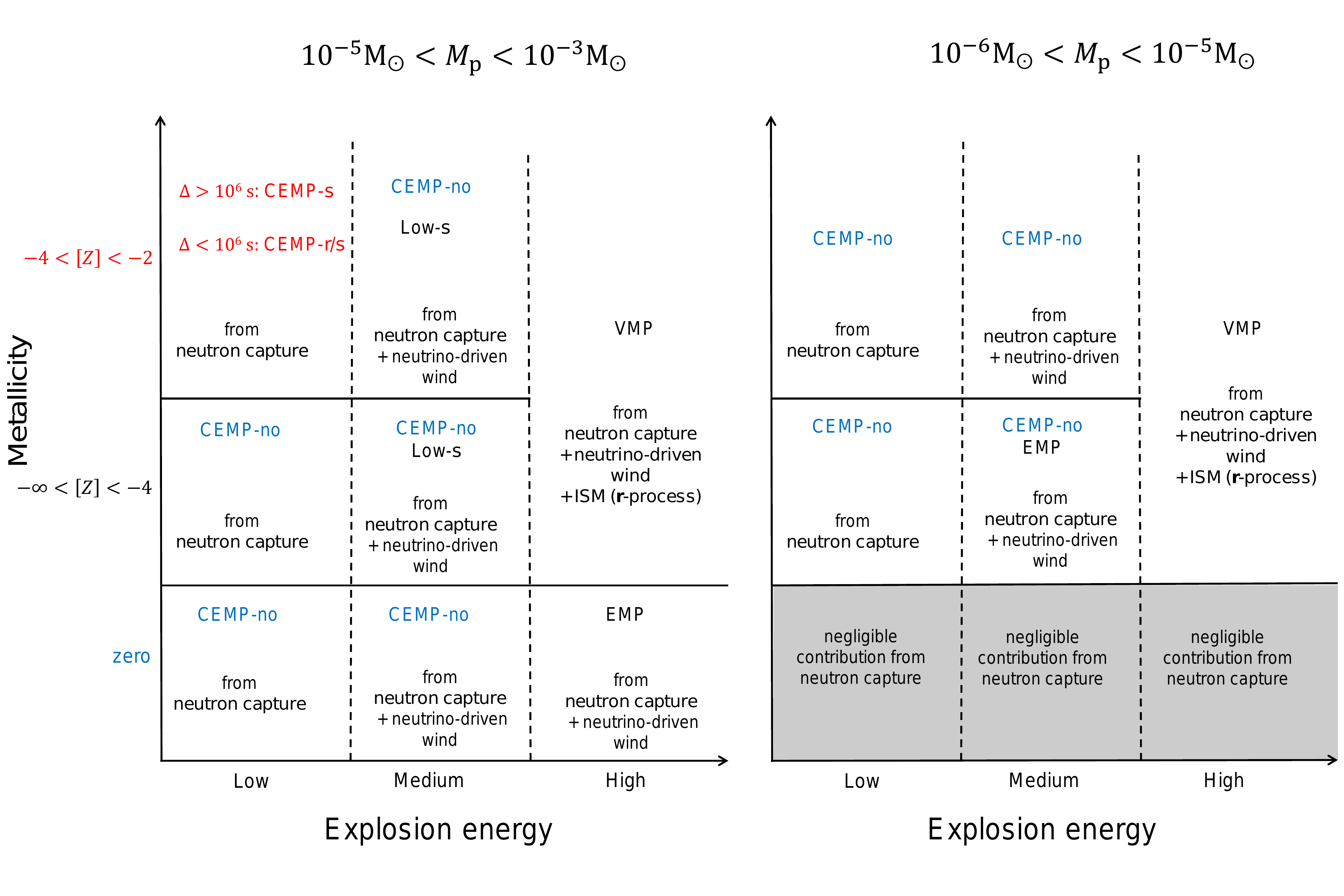}}
\caption{Schematic framework for relating the proposed neutron-capture site in early massive stars to
various classes of VMP stars. Three sources for heavy elements are indicated: neutron
capture at the proposed site, neutrino-driven winds from the associated CCSN, 
and \textsl{r}-process enrichment in the ISM.
VMP and EMP stand for very ($[\textrm{Fe}/\textrm{H}]\lesssim -2$) and extremely ($[\textrm{Fe}/\textrm{H}]\lesssim -3$)
metal-poor stars, respectively. Low-\textsl{s} stands for stars with \textsl{s}-like patterns but low abundances of
heavy elements.}
\label{fig:schematic}
\end{figure*}

With the heavy-element yields proportional to the metallicity for the proposed site, CCSNe from
massive stars with $-4\lesssim[\textrm{Fe}/\textrm{H}]\lesssim -2$ would result in substantially 
higher enrichment. The level of enrichment 
also depends on the amount of dilution and hence on 
the CCSN explosion energy $E_{\rm expl}$. 
CCSNe with low to medium $E_{\rm expl}$ can produce most of the range 
$0\lesssim \log \epsilon({\rm Ba})\lesssim 2.5$ observed in CEMP-\textsl{s} and 
CEMP-\textrm{r}/\textrm{s} stars.  For example, $\log\epsilon({\rm Ba})\sim 2$ can 
be obtained from $25\,\mathrm{M}_\odot$ models with $[\textrm{Fe}/\textrm{H}]\sim -3$ 
and $10^{-5}\,\mathrm{M}_\odot\lesssim M_\mathrm{p}\lesssim 10^{-3}\,\mathrm{M}_\odot$ for a dilution mass of 
$\sim 10^2\,\mathrm{M}_\odot$ corresponding to low $E_{\rm expl}$. CCSNe with higher $E_{\rm expl}$
can explain CEMP-\textsl{s}, CEMP-no, or even normal non-CEMP stars, depending on the 
details of mixing among the heavy elements, C, and Fe, all
of which are produced in distinct regions. 
In general, CCSNe with low to medium $E_{\rm expl}$
would eject most of the material and hence the heavy elements in the He shell, along with a 
larger fraction of C than Fe in the inner part. Such CCSNe would account for
CEMP stars with $[\textrm{C}/\textrm{Fe}]>0.7$ and $[\textrm{Ba}/\textrm{Fe}]$ from 
$<0$ (CEMP-no) to $>1$ (CEMP-\textsl{s}), as well as VMP stars with $0<[\textrm{Ba}/\textrm{Fe}]<1$. 
The characteristic $[\textrm{Ba}/\textrm{Eu}]$ for CEMP-\textsl{s}
($> 0.5$) and CEMP-\textrm{r}/\textrm{s} ($\lesssim 0.5$) stars can also be obtained for different
duration $\Delta$ of neutron capture in the pre-CCSN star. The above framework for
relating our proposed neutron-capture site in early massive stars to the diversity of
VMP stars is illustrated in Fig.~\ref{fig:schematic}.

The four VMP stars shown in Fig.~\ref{fig:obs} are difficult to explain by the existing paradigm. 
CS 30301-015 \citep{aoki2002} (Fig.~\ref{fig:obs}a) appears to be a single CEMP-\textsl{s} star \citep{hansenCEMPs} in contradiction 
to the prevalent explanation of CEMP-\textsl{s} stars by mass transfer from an AGB
companion during binary evolution. CS 22964-161 \citep{thompson2008} (Fig.~\ref{fig:obs}b) represents two
CEMP-\textsl{s} stars orbiting each other with no sign of 
a white dwarf left behind by a former AGB companion. Unless the evolution of a triple-star system somehow
resulted in the ejection of the white dwarf and the formation of a binary system \citep{don1992}, 
observations of CS 22964-161 and its current companion are also in conflict with the scenario of surface pollution
by mass transfer. 
In contrast, low-mass CEMP-\textsl{s} stars in both single and binary configurations can form naturally 
from an ISM enriched by a metal-poor CCSN based on our models with 
$\Delta>10^6\,$s.
The same is also true for CEMP-\textsl{r}/\textsl{s} stars based on our models with
$\Delta \lesssim 10^6\,$s. HE 2258-6358 \citep{placco2013} (Fig.~\ref{fig:obs}c) is an example of a 
CEMP-\textsl{r}/\textsl{s} star in good agreement with our model. Finally,
CS 29493-090 \citep{spite2014} (Fig.~\ref{fig:obs}d)
is a CEMP star ($[\textrm{C}/\textrm{Fe}]=0.73$) that does not show any sign of binarity and has an \textsl{s}-like pattern with
a low enrichment of $\log\epsilon({\rm Ba})=-0.49$.
Its intermediate value of $[\textrm{Ba}/\textrm{Fe}]=0.43$ excludes it from being classified as
either a CEMP-no or CEMP-\textsl{s} star.
This and other similar stars reported in \cite{spite2014} 
are difficult to explain by the existing paradigm. 
Such stars with $0<[\textrm{Ba}/\textrm{Fe}]<1$, however, are a natural 
consequence of neutron-capture nucleosynthesis by our proposed site followed by a CCSN 
with low to medium explosion energy as discussed above.

\subsection{{\rm C} Abundances in CEMP Stars}

The majority of CEMP-no stars have lower C enrichment with a mean abundance of $\log \epsilon({\rm C})\sim 6.3 \pm 0.5$
compared to the majority of CEMP-\textsl{s} and CEMP-\textsl{r}/\textsl{s} stars with a mean abundance of $\log \epsilon({\rm C})\sim 8.0 \pm 0.4$ 
\citep{spite2013,bonifacio2015,hansen2015,yoonCEMPno1}. 
This difference in C enrichment is consistent with the framework based on our proposed neutron-capture site 
as illustrated in Fig.~\ref{fig:schematic},
where CCSN sources for CEMP-no stars are more energetic than those for CEMP-\textsl{s} and CEMP-\textsl{r}/\textsl{s} stars.
More energetic CCSNe would have a larger mass of ISM to dilute their ejecta, thereby causing
the lower C enrichment for CEMP-no stars. For example, a $25 \,\mathrm{M}_\odot$ star produces $0.3 \,\mathrm{M}_\odot $ of C in total, which
must be diluted by $\sim 1.6\times 10^4 \,\mathrm{M}_\odot$ of ISM to give $\log \epsilon({\rm C})\sim 6.3$.
This dilution mass is expected for a CCSN with an explosion energy of $\sim 10^{51}\,$ergs. 
Such a source with $[Z]\lesssim -2$ would also result in 
$\log \epsilon({\rm Ba})\lesssim 0$, similar to the Ba enrichment observed in many CEMP-no stars. 
On the other hand, a low dilution of $\lesssim 1,\!000 \,\mathrm{M}_\odot$ as required to explain
CEMP-\textsl{s} and CEMP-\textsl{r}/\textsl{s} stars would correspond to a faint CCSN
with substantial fallback of the inner material including the C and O shells. 
The presence of $\sim 0.1 \,\mathrm{M}_\odot$ of C in the He shell, however,  can still result in $\log \epsilon({\rm C})\sim 7$--$8$ 
for a dilution mass of $\sim 100$--$1,\!000 \,\mathrm{M}_\odot$. This estimate is consistent with the C enrichment found in CEMP-\textsl{s} 
and CEMP-\textsl{r}/\textsl{s} stars. Ejection of additional C and O from the inner C and O shells would depend on the details of mixing.

\section{Summary and Outlook}
\label{sec-sum}

We have proposed a new neutron-capture site associated with proton ingestion in
convective He shells of early stars of $\sim 20$--$30\,\mathrm{M}_\odot$.
The nucleosynthesis by this site has similar characteristics
to the \textsl{i}-process that has been studied in connection with proton ingestion in
AGB stars. We find that although an \textsl{r}/\textsl{s}-like pattern typically results from
the \textsl{i}-process with high exposure at high neutron densities, this pattern
can be easily and quickly changed to an \textsl{s}-like pattern by subsequent low
exposure at low neutron densities. Thus, attaining \textsl{i}-process neutron densities
does not necessarily result in \textsl{r}/\textsl{s}-like patterns. To retain such patterns requires
no subsequent low exposure at low neutron densities. This requirement can be
fulfilled by the onset of core collapse of the massive star associated with our
proposed site. On the other hand, it is difficult to avoid some low 
exposure at low neutron densities following the \textsl{i}-process in AGB stars.
How this exposure may modify the yield pattern 
of the \textsl{i}-process remains to be investigated.

As discussed in \S\ref{sec-discuss}, our proposed site can potentially account for 
both single and binary CEMP stars with a wide range of patterns and absolute abundances 
of heavy elements. Consequently, at least some CEMP-\textsl{s} and CEMP-\textsl{r}/\textsl{s} stars 
may have originated directly from the ISM in the same way as CEMP-no stars 
(see Fig.~\ref{fig:schematic}). CEMP-\textsl{s} and CEMP-\textsl{r}/\textsl{s} stars require CCSN
sources with low energy explosions to ensure low dilution of nucleosynthesis
products from the proposed site, whereas CEMP-no stars can have sources with
medium energy explosions. This scenario is possible because the heavy elements 
synthesized in the He shell can be ejected even by weak explosions.

In the current paradigm, only CEMP-no stars are thought to represent 
the composition of the early ISM enriched by the very first massive stars.
The widely accepted scenario for the formation of a CEMP-\textsl{s} star is surface pollution 
through binary mass transfer from a heavier companion during its AGB phase.
In this scenario, the CEMP-\textsl{s} star should be observed today to orbit around
the white dwarf left behind by the former AGB companion, which is consistent with 
the high binary frequency of $\sim 80\,\%$ for such stars as compared to 
$\sim 17\,\%$ for CEMP-no stars \citep{hansenCEMPno,hansenCEMPs}.
In addition, this scenario can explain both the abundance pattern 
and the large enhancement of heavy elements and C for most CEMP-\textsl{s} stars. 
The $\sim 20\,\%$ of CEMP-\textsl{s} stars that appear to be single, however, require
that their surface abundances reflect the composition of the ISM from which they formed
instead of pollution by binary mass transfer. Enrichment of the ISM by CCSNe associated
with our proposed site would account for these single CEMP-\textsl{s} stars (see Fig.~\ref{fig:schematic}).

We note that spinstars may also be sources for single CEMP-\textsl{s} stars \citep{choplin2017}. 
In fact, this model can explain the four apparent single CEMP-\textsl{s} stars found by \citet{hansenCEMPs}
reasonably well. Its potential weakness is that it can only account for very little Pb.
Incidentally, three of the above four stars do not have Pb measurement yet, and most CEMP-\textsl{s} stars 
with Pb measurement tend to have high Pb abundances. Future measurement of Pb abundances in single
CEMP-\textsl{s} stars would help clarify whether spinstars can also be sources for such stars.

Because stars in both single and binary configurations can form from the ISM,
some CEMP-\textsl{s} stars in binaries could have formed from the same ISM that gave
birth to single low-mass CEMP-\textsl{s} stars. 
If a binary of two low-mass stars was formed, this scenario would explain binaries of 
two CEMP-\textsl{s} stars like CS~22964-161 \citep{thompson2008}.
On the other hand, if a low-mass star was formed with a companion of 
$\sim 1$--$8 \,\mathrm{M}_\odot$, the companion during its AGB phase would have polluted  
the surface of the low-mass star and subsequently left behind a white dwarf.
This scenario is similar to the current paradigm for CEMP-\textsl{s} stars, except that the low-mass
star was born with heavy elements and its birth composition was later superposed on its surface 
with the new C and \textsl{s}-process products from its AGB companion. This interesting scenario may
lead to some special features in the heavy-element pattern and should be explored by
further studies. 

Although the origin of CEMP-\textsl{r}/\textsl{s} stars is still uncertain, all the previously 
proposed scenarios are based on surface pollution by binary mass transfer from a companion of 
low to intermediate mass, similarly to CEMP-\textsl{s} stars.
The binary fraction of CEMP-\textsl{r}/\textsl{s} stars is unclear at present due to limited data. Determination of
this fraction by future surveys is particularly important in evaluating our proposed 
neutron-capture site as an explanation for CEMP-\textsl{r}/\textsl{s} stars. 
Our framework based on this site would predict that 
single CEMP-\textsl{r}/\textsl{s} stars should occur with a frequency comparable to single CEMP-\textsl{s} stars.
In addition, there should be ``low-\textsl{s}'' stars with similar patterns of heavy elements to 
CEMP-\textsl{s} and CEMP-\textsl{r}/\textsl{s} stars but with low enrichment of these elements ($0<[\textrm{Ba}/\textrm{Fe}]<1$).
CS 29493-090 \citep{spite2014} is such an example. In general,
the low-\textsl{s} stars should have similar C enhancement to CEMP-no stars and
be observed both as single stars and with a lower binary frequency than CEMP-\textsl{s} stars. 

At $[\textrm{Fe}/\textrm{H}]\gtrsim -3.5$, the ISM could have been polluted by \textsl{r}-process  
sources such as neutron star mergers and jet-driven CCSNe.
If neutron-capture nucleosynthesis by our proposed site in massive stars of those times 
was followed by CCSNe with high explosion energies of 
$\gtrsim 10^{51}\,$ergs, heavy elements produced with an \textsl{s}-like or \textsl{r}/\textsl{s}-like pattern would be mixed 
broadly with the \textsl{r}-process products in the ISM. Stars formed from such an ISM would have patterns 
with clear deviation from the \textsl{r}-process kind. Indeed, VMP stars with $[\textrm{Fe}/\textrm{H}]$ as low as $\sim -3.5$ were 
observed to have non-\textsl{r}-process values of $[\textrm{Ba}/\textrm{Eu}]$ and $[\textrm{La}/\textrm{Eu}]$ \citep{simmerer,saga}. 
This observation is consistent with the above scenario, but difficult to explain otherwise because 
no pervasive \textsl{s}-process enrichment of the ISM is expected from AGB stars at such early times.
We predict that future measurements of detailed heavy-element patterns in single VMP stars will reveal 
many cases of mixed contributions from the \textsl{r}-process and the neutron-capture nucleosynthesis by our 
proposed site. We note that HD~94028 with [Fe/H]~$\approx-1.6$, somewhat outside the range of [Fe/H]~$\lesssim -2$
for the VMP stars of concern here, appears to have received neutron-capture elements from multiple sources \citep{roederer2016}.
Whereas we have focused on Sr, Ba, La, Eu, and Pb as specific elements of interest, the element As in this and perhaps other
similar stars may have originated from the \textsl{i}-process \citep{roederer2016}. This important issue and the overall patterns of
neutron-capture elements in such stars are outside our scope here but certainly merit further detailed studies.

With the proposed site operating in metal-free and metal-poor stars of $\sim 20$--$30\,\mathrm{M}_\odot$, 
important constraints may be set on the formation mode for stars of populations III and II. 
The present study has focused on the heavy elements. In the future, we plan to include both these and 
lighter elements in the mixing and fallback prescription for CCSN ejecta 
to make full comparison with observations,
thereby probing the allowed mass range of early CCSN sources. We note that other recent studies 
used the patterns of elements up to the Fe group in CEMP-no stars to constrain 
the mass range of their metal-free CCSN sources \citep{placco2016}. How the allowed mass range
would be affected by including the heavy elements from our proposed site in such studies 
remains to be seen.

Several crucial aspects of our proposed neutron-capture site require further investigation. 
The foremost is accurate 3D modeling of proton ingestion and especially 
the subsequent mixing in He shells of metal-free and metal-poor massive stars. 
Neutron production at this site relies on successful transport of the ingested protons to the base 
of the He shell. This transport can be modelled properly only in 3D. Finally, reliable simulations of 
the associated CCSNe are also important, because the ejection of various nucleosynthesis products from 
different layers is crucial in determining the overall yield pattern including both the heavy 
and lighter elements for comparison with observations. 

\acknowledgements
We thank the anonymous referee for criticisms and suggestions that helped improve the paper.
We made extensive use of the SAGA database \citep{saga}.
This work was supported in part by the NSFC [11533006 (SJTU)], the US DOE 
[DE-FG02-87ER40328 (UM)], and the ARC [FT120100363 (Monash)]. 

\software{\textsc{Kepler}} \citep{weaver1978, rauscher2003}



\begin{table*}
\centering
\vspace{-\baselineskip}
\caption{References for key reaction rates. BAAL: \citet{BAAL}, W94: \citet{W94}, FCZ2: \citet{FCZ2}, HFCZ: \citet{HFCZ}, 
BU96: \citet{BU96}, CF88: \citet{CF88}, WIES: \citet{WIES}, RATH: \citet{RATH}, FKTH: \citet{FKTH1,FKTH2}, RA94: \citet{RA94}, 
LA90: \citet{LA90}, HW01: \citet{HW01}.}
\vskip 0.5cm
\begin{tabular}{llllllllll}
\hline
$^AZ$       & $(\mathrm{n},\gamma)$  &$(\mathrm{n},\mathrm{p})$ & $(\mathrm{n},\alpha)$ & $(\mathrm{p},\mathrm{n})$ & $(\mathrm{p},\gamma)$ & $(\mathrm{p},\alpha)$& $(\alpha,\mathrm{p})$  & $(\alpha,\mathrm{n})$  & $(\alpha,\gamma)$ \\
\hline
$^{12}$C    &BAAL           &        &W94           &         &FCZ2          &             &HFCZ           &HFCZ           &BU96\\
$^{13}$C    &RA94           &        &              &FCZ2     &CF88          &             &               &FCZ2           &     \\
$^{13}$N    &WIES           &FCZ2    &CF88          &         &KL93          &             &HFCZ           &               &     \\
$^{14}$N    &WIES           &CF88    &CF88          &FCZ2     &CF88          &CF88         &HW01           &CF88           &FCZ2\\
$^{16}$O    &BAAL           &        &FCZ2          &         &FCZ2          &HFCZ         &HFCZ           &RATH           &CF88\\
$^{17}$O    &RA94           &        &              &FKTH     &LA90          &HW01         &               &FCZ2           &FCZ2\\
\hline
\end{tabular}
\vspace{-0.8\baselineskip}
\label{tab:rate}
\end{table*}

\begin{table*}
\centering
\vspace{-\baselineskip}
\caption{Pre-CCSN yields (in $\,\mathrm{M}_\odot$) of Sr, Ba, La, Eu, and Pb 
for $25 \,\mathrm{M}_\odot$ stars with metallicity $[Z]$ and ingested proton mass $M_\mathrm{p}$.
Ingestion occurs only once at central C depletion unless noted otherwise.
Here $X(Y) \equiv X\times10^Y$.}
\vskip 0.5cm
\begin{tabular}{lllllll}
\hline
 Model                                                  &Sr         &Ba     &La           &Eu       &Pb \\
\hline
$[Z]=-\infty, M_\mathrm{p}=10^{-3} \,\mathrm{M}_\odot$                      & 9.85(-14)&1.16(-14) &3.21(-16)& 2.93(-17) &6.74(-17) \\
$[Z]=-\infty$, 3 pulses  of $M_\mathrm{p}=10^{-4} \,\mathrm{M}_\odot$
 every $10^6\,$s                                        & 4.65(-11)&1.44(-10) &6.40(-12)& 2.96(-13) &3.09(-10)\\
$[Z]=-\infty$, 2 pulses  of $M_\mathrm{p}=10^{-4} \,\mathrm{M}_\odot$
 every $10^6\,$s                                        & 1.69(-11)&5.91(-11) &2.72(-12)& 1.26(-13) &1.09(-10)\\
$[Z]=-\infty, M_\mathrm{p}=10^{-4} \,\mathrm{M}_\odot$                      & 6.42(-12)&1.85(-11) &8.17(-13)& 3.37(-14) &1.67(-11)\\
$[Z]=-\infty, M_\mathrm{p}=5\times10^{-5} \,\mathrm{M}_\odot$              & 5.12(-12)&1.39(-11) &6.08(-13)& 2.35(-14) &1.02(-11)\\
$[Z]=-\infty, M_\mathrm{p}=10^{-5} \,\mathrm{M}_\odot$                      & 5.80(-13)&2.70(-13) &8.76(-15)& 1.96(-16) &1.45(-14)\\
$[Z]=-\infty, M_\mathrm{p}=10^{-6} \,\mathrm{M}_\odot$                      & 1.82(-14)&2.63(-15) &1.71(-16)& 2.51(-17) &0.00  \\
\hline
$[Z]=-5, M_\mathrm{p}=10^{-3} \,\mathrm{M}_\odot$                           & 2.41(-9) &2.15(-9)  &8.87(-11)& 4.01(-12) &5.92(-10)\\
$[Z]=-5, M_\mathrm{p}=10^{-4} \,\mathrm{M}_\odot$                           &9.21(-10) &4.71(-9)  &2.66(-10)&1.67(-11)  &2.09(-8)\\
$[Z]=-5, M_\mathrm{p}=10^{-5} \,\mathrm{M}_\odot$                           & 2.28(-9) &4.55(-9)  &1.84(-10)&7.80(-12)  &1.31(-9) \\
$[Z]=-5, M_\mathrm{p}=10^{-6} \,\mathrm{M}_\odot$                           & 2.88(-9) &7.78(-10) &3.70(-11)& 1.17(-12) &3.70(-11)\\
\hline

$[Z]=-4, M_\mathrm{p}=10^{-3} \,\mathrm{M}_\odot$                           & 8.90(-9) &5.32(-8)  &2.76(-9)  & 1.57(-10) &1.51(-7)\\  
$[Z]=-4, M_\mathrm{p}=10^{-4} \,\mathrm{M}_\odot$                           & 9.84(-9) &5.56(-8)  &3.07(-9)  & 1.46(-10) &1.64(-7) \\                   
$[Z]=-4, M_\mathrm{p}=10^{-5} \,\mathrm{M}_\odot$                           &1.84(-8)  &1.10(-8)  &3.90(-10)& 6.67(-12)  &4.09(-10) \\   
$[Z]=-4, M_\mathrm{p}=10^{-6} \,\mathrm{M}_\odot$                           &2.81(-10)&1.66(-14)&1.59(-16) &2.41(-18)    &0.00    \\          
                                                   
\hline
$[Z]=-3, M_\mathrm{p}=10^{-3} \,\mathrm{M}_\odot$                           & 8.74(-8)  &5.22(-7)  &2.66(-8) & 1.32(-9) &1.38(-6) \\
$[Z]=-3, M_\mathrm{p}=10^{-4} \,\mathrm{M}_\odot$                           & 9.49(-8)  &5.44(-7)  &3.03(-8)& 1.26(-9)  &1.63(-6)\\
$[Z]=-3, M_\mathrm{p}=5\times10^{-5} \,\mathrm{M}_\odot$                    & 1.21(-7)  &5.73(-7)  &3.12(-8)& 1.25(-9)  &9.53(-7)\\
$[Z]=-3, M_\mathrm{p}=10^{-5} \,\mathrm{M}_\odot$                           & 1.86(-7)  &1.42(-7)  &5.70(-9)& 1.52(-10) &8.60(-9)\\
$[Z]=-3, M_\mathrm{p}=5\times10^{-6} \,\mathrm{M}_\odot$                    & 1.03(-7)  &1.19(-8)  &3.49(-10)& 6.85(-12)&5.38(-11)\\
$[Z]=-3, M_\mathrm{p}=10^{-6} \,\mathrm{M}_\odot$                           & 2.23(-9)  &2.24(-13) &1.66(-15)&3.71(-18) &0.00   \\
\hline
$[Z]=-2, M_\mathrm{p}=10^{-3} \,\mathrm{M}_\odot$                           & 8.37(-7)  &4.92(-6)  &2.41(-7)& 1.43(-8)&1.26(-5)\\
$[Z]=-2, M_\mathrm{p}=10^{-4} \,\mathrm{M}_\odot$                           & 1.03(-6)  &4.73(-6)  &2.23(-7)& 1.37(-8)  &7.47(-6)    \\
$[Z]=-2, M_\mathrm{p}=10^{-5} \,\mathrm{M}_\odot$                           & 2.15(-7)  &2.84(-9)  &3.34(-11)& 3.78(-13)&6.55(-13)\\
$[Z]=-2, M_\mathrm{p}=10^{-6} \,\mathrm{M}_\odot$                           & 3.02(-12) &0.00      &0.00     & 0.00     &0.00   \\
\hline
$[Z]=-1, M_\mathrm{p}=10^{-3} \,\mathrm{M}_\odot$                           & 1.19(-5)  &1.76(-5)  &6.06(-7)& 2.11(-8)&3.80(-6)\\
$[Z]=-1, M_\mathrm{p}=10^{-4} \,\mathrm{M}_\odot$                           & 5.23(-5)  &2.66(-7)  &7.70(-9)& 6.34(-10)&5.15(-8)\\
$[Z]=-1, M_\mathrm{p}=5\times 10^{-5} \,\mathrm{M}_\odot$                   & 8.39(-7)  &5.86(-8)  &4.77(-9)& 5.93(-10)&4.94(-8)\\
$[Z]=-1, M_\mathrm{p}=10^{-5} \,\mathrm{M}_\odot$                           & 3.86(-7)  &5.79(-8)  &4.79(-9)& 5.93(-10)&4.63(-8)\\
$[Z]=-1$, without proton ingestion                      & 3.87(-7)  &5.69(-8)  &4.70(-9)& 5.94(-10)&4.51(-8)\\
\hline
\end{tabular}
\vspace{-0.8\baselineskip}
\label{tab:1}
\end{table*}

\begin{table*}
\centering
\vspace{-\baselineskip}
\caption{Post-CCSN yields (in $\,\mathrm{M}_\odot$) of Sr, Ba, La, Eu, and Pb 
for $25 \,\mathrm{M}_\odot$ stars with metallicity $[Z]$, ingested proton mass $M_\mathrm{p}$, and explosion energy $E_{\rm expl}$
(Models 1--6) used for comparison with VMP stars. Here $X(Y) \equiv X\times10^Y$, and Cdep (Odep) denotes
central C (O) depletion.}
\vskip 0.5cm
\begin{tabular}{lllllll}
\hline
 Model                                                                          &Sr       &Ba         &La       &Eu        &Pb \\
\hline
Model 1:  $[Z]=-3, M_\mathrm{p}=5\times10^{-5} \,\mathrm{M}_\odot$ at Cdep, $E_{\rm expl}=10^{50}\,$ergs&1.21(-7)  &5.75(-7)   &3.29(-8) & 1.51(-9) &9.58(-7)\\
Model 2:  $[Z]=-3, M_\mathrm{p}=10^{-4} \,\mathrm{M}_\odot$ at Cdep, $E_{\rm expl}=10^{50}\,$ergs        &9.48(-8) &5.45(-7)   &3.16(-8) & 1.50(-9) &1.64(-6)\\
Model 3:  $[Z]=-3, M_\mathrm{p}=10^{-4} \,\mathrm{M}_\odot$ at Odep, $E_{\rm expl}=10^{50}\,$ergs        & 8.52(-8)&4.12(-7)   &2.52(-8) & 2.69(-9) &9.93(-7)\\
Model 4: $[Z]=-4, M_\mathrm{p}=10^{-4} \,\mathrm{M}_\odot$ at Odep, $E_{\rm expl}=3\times10^{50}\,$ergs  &8.57(-9) &4.15(-8)   &2.56(-9)& 2.82(-10) &1.03(-7)\\
Model 5: $[Z]=-2.5$, 2 pulses  of $M_\mathrm{p}=5\times 10^{-4} \,\mathrm{M}_\odot$ at Cdep         & 1.84(-7)&1.71(-6)   &1.09(-7)& 7.50(-9) &7.48(-6)\\
\phantom{Model 5:} and $10^6\,$s later, $E_{\rm expl}=10^{50}\,$ergs                                         &&&&&\\   
Model 6: $[Z]=-2.5$, 2 pulses  of $M_\mathrm{p}=5\times10^{-4} \,\mathrm{M}_\odot$  at Cdep         & 1.91(-7)&1.35(-6)   &9.50(-8)& 1.19(-8) &6.06(-6)\\
\phantom{Model 6:} and Odep, $E_{\rm expl}=10^{50}\,$ergs                                               &&&&&\\   
\hline
\end{tabular}
\vspace{-0.8\baselineskip}
\label{tab:2}
\end{table*}

\end{document}